\documentstyle[12pt]{article}
\begin{document}

\begin{titlepage}
\begin{flushright}
DPSU-95-7\\
August 1995
\end{flushright}

\vskip 2cm

\begin{center}
{\Large Low-Energy Effective Lagrangian}

\vskip 0.2cm

{\Large from Non-Minimal Supergravity}

\vskip 0.2cm

{\Large with Unified Gauge Symmetry}

\vskip 2cm

{\Large Yoshiharu~Kawamura}

\vskip 0.5cm

{\large\sl Department of Physics, Shinshu University\\
Matsumoto, 390 Japan}
\end{center}

\vskip 2cm

\begin{center}
{\bf ABSTRACT}
\end{center}
{}From general supergravity theory with unified gauge symmetry,
we obtain the low-energy effective Lagrangian
by taking the flat limit and integrating out the
superheavy fields in model-independent manner.
The scalar potential possesses some excellent features.
Some light fields classified by using supersymmetric fermion
mass, in general, would get intermediate masses at the tree
level after the supersymmetry is broken.
We show that the stability of weak scale can be guaranteed
under some conditions.
There exist extra non-universal contributions
to soft supersymmetry breaking terms
which can give an impact on phenomenological study.

\end{titlepage}

\newpage

\section{Introduction}

The minimal supersymmetric standard model (MSSM) is the
most attractive candidates for the realistic theory
beyond standard model.
The naturalness problem is elegantly solved by the introduction
 of supersymmetry (SUSY)\cite{Naturalness}.
The SUSY requires new particles called $\lq$superpartners'.
Those masses are free parameters in the MSSM,\footnote{
In this paper, we do not assume the universality on the
soft SUSY breaking parameters from the beginning
 when we use the terminology $\lq$MSSM'.} but are estimated
as at most order 1 TeV from the naturalness argument.
The search for $\lq$superpartners' is one of the main purpose
in the experimental projects by the use of huge colliders,
which have been planed now\cite{JLC}.
% to resolve the mechanism of the electroweak symmetry breaking
%and to discover new physics at TeV region.

It is, however, believed that the MSSM is not the ultimate theory
because there are many problems not to be solved in it.
Here we pick up two problems.
First there are so many free parameters
to be fixed only by experiments for the present.
In addition to gauge couplings and Yukawa couplings,
 soft SUSY breaking parameters appear, i.e., gaugino masses,
 scalar masses and scalar trilinear couplings are
all arbitrary ones.
Hence the MSSM lacks predictability.
Second the mechanism of SUSY breaking is unexplained.
This problem is partly related to the first one since
 the pattern of soft SUSY breaking terms
 depends on the SUSY breaking mechanism.

It is expected that they are solved in more fundamental
 theory.
Supergravity theory (SUGRA)\cite{SUGRA} is
an attractive candidate.
When we take SUGRA as an effective theory
at the Planck scale $M_{Pl}$,
SUGRA has an interesting solution.
There exists such a scenario\cite{Hidden}
that the SUSY is spontaneously broken
in the so-called hidden sector
and the effect is transported to the observable sector
through the gravitational interaction.
As a result, the soft SUSY breaking terms appear
in our visible sector.
The form of soft SUSY breaking terms
is determined by the structure of SUGRA.
A simple choice is a theory such that
 soft parameters take universal values at the
gravitational scale $M \equiv M_{Pl}/{\sqrt {8\pi}}$, e.g.,
the scalar potential derived from the SUGRA
with a minimal K\"ahler function has
the universal scalar mass $m_0$ and
 the universal scalar trilinear coupling constant $A$.
% and the universal mixing mass parameter $B$ given by $B=A-m_0$.
%(The universal gaugino mass $M_{1/2}$ is derived from SUGRA with
%the gauge kinetic function $f_{\alpha \beta}
%= S\delta_{\alpha \beta}$.)
Those values at low energy are calculated by using
renormalization group equations.

The analyses based on the MSSM are energetically
investigated\cite{MSSM}.
Most of them are highly constrained by the assumption
that the soft SUSY breaking parameters are universal at $M$
or a unification scale $M_X$.
This assumption is quite interesting
because the theory has high predictabilities
and is testable enough, but it is difficult to say that this
type of approach is completely realistic.
Let us describe the reason why the universality at $M$ is
not necessarily realistic.
First the assumption of the universal scalar mass is motivated
by the fact that the flavor-changing neutral current
(FCNC) processes are suppressed experimentally\cite{FCNC}.
However, we can relax this assumption since the suppression
of FCNC processes due to SUSY particle loops requires only
the degeneracy among squarks with a same flavor.
Second there is no strong reason that the realistic SUGRA takes
the minimal structure.
In fact, the effective SUGRAs derived from superstring theories
(SSTs) have, in general, non-minimal structures and they can
lead to the effective theories
with non-universal soft parameters\cite{SST}.
Third it was pointed out that higher order corrections generally
destroy the minimal form of the K\"ahler potential\cite{Rad-cor}.
Last the effects of supersymmetric grand unified theory (SUSY-GUT)
were little considered in the analysis of the running of
parameters although SUSY-GUT\cite{SUSY-GUT} has been hopeful
as a realistic theory.

We shall discuss the last point still more.
The unification dogma\cite{GUT} has a merit that the number of
independent parameters is reduced due to a large gauge symmetry.
Further SUSY $SU(5)$ GUT is supported by the LEP
experiments\cite{LEP} and predicts the long lifetime of nucleon
consistent with the present data\cite{Decay}.
Thus an analysis based on the MSSM encouraged by
the unification scenario seems to be hopeful.
In fact, many researches have been done under the assumption
 that the soft SUSY breaking terms take a universal form
at the GUT scale $M_X$, but this assumption is also not
always realistic from the following reasons.
The non-minimal SUGRA can lead to the non-universal form
of soft SUSY breaking terms as described the above.
Even if we take a minimal SUGRA as a starting point,
 the radiative correction from $M$ to $M_X$ changes
the universal form of SUSY breaking terms into non-universal one.
In some literatures, the renormalization effects
were discussed\cite{Ren-eff}, but we need to consider
effects on the gauge symmetry breaking further.
In Ref.~\cite{KMY2}, low-energy effective theory has
been derived from SUSY-GUT with non-universal soft SUSY
breaking terms by integrating out superheavy fields
and it is shown that new contributions to SUSY breaking terms
can appear at the tree level
after the breaking of unified gauge symmetry.
The analyses including the effects are started
recently\cite{Re-analyses}.

Now we should stress the importance of studying
the soft SUSY breaking terms.
The reason is that they can be a powerful probe to SUSY-GUT
 and/or SUGRA since the weak scale SUSY spectrum
can directly reflect the physics at very high energies.
For example, we can check the GUT scenario experimentally
by measuring the gaugino masses\cite{Gaugino}.
Also, the scalar mass spectrum has certain ``sum rules''
specific to symmetry breaking patterns\cite{Sfermion}.
Therefore, the precision measurements of SUSY spectrum
are very important.
And it is a meaningful subject to obtain the low-energy theory
in more general framework and to grasp the peculiarities
concerning on the SUSY breaking terms in advance.

Various types of low-energy theories were derived
 as will be explained in the next section.
However its low-energy theory has not been completely
investigated by taking SUGRA with general structure and
unified gauge symmetry as a starting point.
It has been only studied
in some specific cases\cite{HLW}\cite{Drees}\cite{KMY2}.
For example, it is shown that the universality of scalar masses
is preserved in the SUGRA whose K\"{a}hler potential
has $U(n)$ symmetry among the $n$ chiral fields\cite{HLW}.
In Ref.\cite{KMY2},
the scalar potential was derived starting from a unified theory
with a certain type of non-universal soft SUSY breaking terms.
Such non-universal soft terms arise if we take the flat limit
of the SUGRA where the K\"{a}hler potential is a certain type
of non-minimal one and the superpotential
 is separate from the hidden sector to the visible one.
(We call this form of superpotential a ${\it hidden}$ ansatz.)

In this paper, we derive the low-energy effective theory from
non-minimal SUGRA with unified gauge symmetry.
The starting SUGRA has more general structure, i.e.,
the K\"{a}hler potential is non-minimal and
we do not impose the ${\it hidden}$ assumption
on the superpotential.
Then dangerous terms, which destabilize the gauge hierarchy,
generally appear at the tree level.
We discuss conditions that the hierarchy is preserved, and
 take the flat limit and integrate out superheavy fields without
identifying $M_X$ with $M$.
We find various contributions to the SUSY breaking terms.
Our result reduces to that obtained in Ref.~\cite{HLW} in
the case with the minimal SUGRA.
Also it is shown that it reduces to that obtained
in Ref.\cite{KMY2} in the limit $M_X/M \to 0$
when we take a certain type of total K\"{a}hler potential.

The paper is organized as follows.
In section 2, we first review the low-energy effective
Lagrangians from SUGRA following the historical development.
We derive the low-energy effective scalar potential
starting from SUGRA with general total K\"{a}hler potential and
 unified gauge symmetry in section 3.
In section 4, we discuss $D$-term contributions to
scalar masses and make clear the relation between our result
and that in Ref.\cite{KMY2}.
Section 5 is devoted to conclusions.

\section{Historical Background}
%\cleqn
\subsection{Scalar Sector in SUGRA}

We begin by reviewing the scalar sector in SUGRA\cite{SUGRA}.
It is specified by two functions,  the total K\"ahler
potential $G(\Phi, \Phi^*)$ and the gauge kinetic function
$f_{\alpha \beta}(\Phi)$ with $\alpha$, $\beta$
being indices of the adjoint representation of the gauge group.
The former is a sum of the K\"ahler potential $K$
and (the logarithm of) the superpotential $W_{SG}$ such as
\begin{eqnarray}
   G(\Phi, \Phi^*)=K(\Phi, \Phi^*)
+M^{2}\ln |W_{SG}(\Phi) /M^{3}|^2.
\label{G}
\end{eqnarray}
We have denoted the chiral multiplets by $\Phi^{I}$
and their complex conjugate by $\Phi_{J}^*$.
The scalar potential is given by
\begin{eqnarray}
   V= M^{2}e^{G/M^{2}} U
     +\frac{1}{2} (Re f^{-1})_{\alpha \beta} D^{\alpha} D^{\beta},
\label{V}
\end{eqnarray}
where
\begin{eqnarray}
  U &=& G^I (K^{-1})_I^J G_{J}-3M^{2},
\label{U}
\\
  D^\alpha &=&  G_I( T^\alpha z)^I
            = (z^\dagger T^\alpha)_J G^J.
\label{D}
\end{eqnarray}
Here $G_{I}=\partial G/\partial z^I$,
$G^{J}=\partial G/\partial z_{J}^*$ etc, and
$T^\alpha$ are gauge transformation generators.
The $z^I$ is a scalar component of $\Phi^{I}$.
Here and hereafter both $G$ and $f_{\alpha \beta}$ are regarded
as functions of $z$ and $z^*$ as we take notice
of the scalar potential alone.
Also $(Re f^{-1})_{\alpha \beta}$ and  $(K^{-1})_I^{J}$
are the inverse matrices of $Re f_{\alpha \beta}$ and  $K_I^{J}$
respectively, and summation over $\alpha$,... and  $I$,... is
understood.  The last equality in Eq.~(\ref{D}) comes from the
gauge invariance of the total K\"ahler potential.

Let us next summarize our assumptions on the SUSY breaking.
The gravitino mass $m_{3/2}$ is given by
\begin{eqnarray}
   m_{3/2} = \langle e^{K/2M^{2}} {W_{SG} \over M^2} \rangle,
\label{m}
\end{eqnarray}
%\begin{eqnarray}
%   m_{3/2}= \langle Me^{G/2M^{2}} \rangle,
%\label{gravitino}
%\end{eqnarray}
where $\langle \cdots \rangle$ denotes
the vacuum expectation value (VEV) of the quantity.
%As a phase convention, it is taken to be real
We identify the gravitino mass with the weak scale.
The $F$-auxiliary fields of the chiral multiplets $\Phi^I$
are defined as
\begin{eqnarray}
   F^I \equiv Me^{G/2M^{2}} (K^{-1})_J^{I} G^J.
\label{F}
\end{eqnarray}
We require those VEVs should satisfy
\begin{eqnarray}
   \langle F^I \rangle \leq O(m_{3/2}M).
\label{<F>}
\end{eqnarray}
We can show that the VEVs of the $D$-auxiliary fields become
 very small $\langle D^\alpha \rangle \leq O(m_{3/2}^2)$
as will be shown in Appendix A.
It follows from
Eqs.~(\ref{U}), (\ref{F}) and (\ref{<F>}) that
\begin{eqnarray}
   \langle G_I \rangle, \ \langle G^J \rangle
   \leq O(M)
\label{<G>}
\end{eqnarray}
and
\begin{eqnarray}
  \langle U \rangle \leq O(M^{2}).
\label{<U>}
\end{eqnarray}
Note that we allow the non-zero
vacuum energy $\langle V \rangle$ of order $m_{3/2}^2 M^2$
at this level, which could be canceled by quantum corrections.
We also assume that derivatives of the K\"ahler potential
with respect to $z$ and $z^*$ are at most of order unity
(in the units where $M$ is taken to be unity), namely
\begin{eqnarray}
  \langle K_{I_1 \cdots}^{J_1 \cdots} \rangle \leq O(1).
\label{KI1J1...}
\end{eqnarray}
This will be justified if the Planck scale physics plays
an essential role in the SUSY breaking.

We shall call the fields which induce to the SUSY breaking
$\lq$hidden fields', and denote those scalar components
and $F$-components as $\tilde{z}^i$
and $\tilde{F}^i$, respectively.
We require those VEVs should satisfy
\begin{eqnarray}
   \langle \tilde{z}^i \rangle &=& O(M)
\label{<zi>}
\end{eqnarray}
and
\begin{eqnarray}
   \langle \tilde{F}^i \rangle &=& O(m_{3/2} M).
\label{<Fi>}
\end{eqnarray}
We shall call the rest $\lq$observable fields' and denote
 the scalar components as $z^\kappa$.

\subsection{Effective Theories from Minimal SUGRA}

The minimal SUGRA is defined as follows.
The K\"ahler potential $K$ has a canonical form as
\begin{eqnarray}
   K = |z^{\kappa}|^2+|\tilde{z}^{i}|^2.
\label{Min-K}
\end{eqnarray}
We take the {\it hidden} ansatz for the superpotential as
\begin{eqnarray}
   W_{SG} = W(z) + \tilde{W}(\tilde{z}).
\label{Hid-W}
\end{eqnarray}

The global SUSY theory with soft SUSY breaking terms
is derived by taking the flat limit,
i.e., $M \to \infty$ but $m_{3/2}$ kept finite.
The scalar potential is as follows\cite{Hidden},
\begin{eqnarray}
V &=& V_{SUSY} + V_{Soft},
\label{Min-V}\\
V_{SUSY} &=& |\frac{\partial \widehat{W}}{\partial z^\kappa}|^2
      + {1 \over 2}g_\alpha^2
(z_\kappa^* (T^\alpha)^\kappa_\lambda z^\lambda)^2 ,
\label{Min-VSUSY}\\
V_{Soft} &=& A \widehat{W}
+ B z^\kappa \frac{\partial \widehat{W}}{\partial z^\kappa}
		+ {\it H.c.}
	+ |B|^2 z_\kappa^* z^\kappa,
\label{Min-Vsoft}
\end{eqnarray}
where $\widehat{W}$ is defined as $\widehat{W}
\equiv \langle exp({K \over 2M^2})\rangle W$.
$V_{SUSY}$ stands for the supersymmetric part, while $V_{Soft}$
 contains the soft SUSY breaking terms.
The $A$ and $B$ are the soft SUSY breaking parameters and
 are written as
\begin{eqnarray}
A &=& {\langle \tilde{F}^i \rangle \langle K_i \rangle \over M^2}
 - 3m_{3/2}^{\ast},
\label{A}\\
B &=& m_{3/2}^{\ast}.
\label{B}
\end{eqnarray}
This form of SUSY breaking terms is referred to as ``universal''.

The low-energy effective Lagrangian derived from the minimal SUGRA
with unified gauge symmetry also has a simple structure.
It was obtained by taking the flat limit and integrating out
superheavy fields simultaneously on the postulation that
 the unification scale $M_X$ is identified with $M$\cite{HLW}.
The low-energy scalar potential takes the following form,
\begin{eqnarray}
V^{eff} &=& V_{SUSY}^{eff} + V_{Soft}^{eff},
\label{Veff}\\
V_{SUSY}^{eff} &=& |\frac{\partial \widehat{W}_{eff}}
{\partial z^k}|^2 + \frac{1}{2} g_a^2 (z_k^* (T^a)^k_l z^l)^2 ,
\label{VeffSUSY}
\\
V_{Soft}^{eff}
	&=& A \widehat{W}_{\it eff}
	+ B z^k \frac{\partial \widehat{W}_{\it eff}}
{\partial z^k }	+ {\it H.c.}
	+ |B|^2 z_k^* z^k + \Delta V,
\label{VeffSoft}
\\
\Delta V &\equiv&  -3A \widehat{W}_{\it eff}
	+ A z^k \frac{\partial \widehat{W}_{\it eff}}
{\partial z^k }	+ {\it H.c.}
\label{DeltaV}
\end{eqnarray}
and it still has the same form as the original one
by a suitable redefinition of the $A$ and $B$ parameters
except the mass squared terms.
Here $z^k$ are the light scalar fields,\footnote{
They assumed that the supersymmetric masses of light fields from
the superpotential are zero.
It is straightforward to generalize their analysis into the case
that the light fields have non-zero but $O(m_{3/2})$ masses.}
 $a$ is the index of generators of unbroken gauge group and
$\widehat{W}_{\it eff}$ is the superpotential $\widehat{W}$
with the extremum values for superheavy fields plugged in.
The scalar mass terms
are still universal with the same mass $B$.\footnote{
Throughout this subsection, it is assumed that the vacuum energy
$\langle V \rangle$ vanishes. In the presence of vacuum energy,
the value of scalar mass $|B|^2$ is replaced by
$|B|^2 + \langle V \rangle/M^2$.}

The universal structure of the low-energy Lagrangian led to
a number of strong conclusions, like the natural absence of
the flavor changing neutral currents \cite{FCNC}
or the radiative breaking scenario due to the heavy top
quark\cite{Rad-br}.
Due to these successes, the phenomenological analysis has
been made in popular based on the SUSY models with
the universal soft SUSY breaking terms\cite{MSSM}.
However, it becomes increasingly apparent that SUGRA
 may not have the minimal form, and it is important to study
the consequences on the low-energy effective Lagrangian.

\subsection{Effective Theories from Non-minimal SUGRA}

The scalar potential is also obtained from the non-minimal SUGRA
with no superheavy fields and the result is given as\cite{S&W},
\begin{eqnarray}
V^{(non)} &=& V_{SUSY}^{(non)} + V_{Soft}^{(non)},
\label{Non-V}\\
V_{SUSY}^{(non)} &=&
\frac{\partial \widehat{\cal W^*}}{\partial z^*_\kappa}
\langle (K^{-1})_\kappa^\lambda \rangle
\frac{\partial \widehat{\cal W}}{\partial z^\lambda}
 + {1 \over 2}g_\alpha^2 (\langle K_\kappa^\lambda \rangle
z_\lambda^* (T^\alpha)^\kappa_\mu z^\mu)^2,
\label{Non-VSUSY}\\
V_{Soft}^{(non)}&=& A \widehat{\cal W}
+ B^\kappa(z)
\langle (K^{-1})_\kappa^\lambda \rangle
\frac{\partial \widehat{\cal W}}{\partial z^\lambda}
		+ {\it H.c.}
\nonumber \\
&~&+ B^\kappa(z)
\langle (K^{-1})_\kappa^\lambda \rangle B_\lambda(z) + C(z, z^*) ,
\label{Non-VSoft}
\end{eqnarray}
where
\begin{eqnarray}
B^\kappa (z) &=& m_{3/2}^*
\langle K^\kappa_\lambda \rangle z^{\lambda}
- K^{\kappa}_j \langle \tilde{F}^j \rangle ,
\label{Bkappa}
\\
C(z, z^*) &=& -\langle \tilde{F}^i \rangle \delta^2 K_i^j
               \langle \tilde{F}^*_j \rangle
\nonumber \\
&~& +\{{1 \over M^2}\langle \tilde{F}^i \rangle
 \langle K_i^j \rangle \langle \tilde{F}^*_j \rangle
 - 3|m_{3/2}|^2\} \delta^2 K
\nonumber \\
&~& + m_{3/2} \langle \tilde{F}^i \rangle \delta^2 K_i + {\it H.c.}
\nonumber \\
&~& - A\{m_{3/2}H(z) -\langle \tilde{F}^*_i \rangle H^i(z) \}
+ {\it H.c.}
\label{C}
\end{eqnarray}
in the case that we take the {\it hidden} ansatz.
Here $\widehat{\cal W}$ is defined as
\begin{eqnarray}
\widehat{\cal W} \equiv \widehat{W} + m_{3/2} H(z)
 - \langle \tilde{F}^*_i \rangle H^i(z),
\label{calW}
\end{eqnarray}
where $H$ is the holomorphic part of $z^{\kappa}$ in $K$.
And $\delta^2 K$, $\delta^2 K_i$ and $\delta^2 K_i^j$ are
the quantities of order $m_{3/2}^2$, $m_{3/2}^2/M$ and
$m_{3/2}^2/M^2$ in $K$, $K_i$ and $K_i^j$, respectively.
Note that the SUSY breaking terms show a non-universal form.
As an excellent feature, there is a natural explanation
for the origin of $\mu$ parameter of order $m_{3/2}$
($\sim$ 1TeV)\cite{G&M}.
That is, the second and third terms in Eq.~(\ref{calW}) correspond
to $\mu$-term with a phenomenologically suitable order.

It is also known that the effective SUGRAs with non-minimal
structure are derived from 4-dimensional string models and
most of them lead to non-universal soft SUSY
breaking terms\cite{SST}.

When the {\it hidden} ansatz is taken off,
the following extra terms should be added,
\begin{eqnarray}
\frac{\partial \widehat{\cal W^*}}{\partial \tilde{z}^*_i}
\langle (K^{-1})_i^j \rangle
\frac{\partial \widehat{\cal W}}{\partial z^j}
 + \Delta C(z, z^*)
+ \langle \tilde{F}^i \rangle
\frac{\partial \widehat{\cal W}}{\partial \tilde{z}^i}
		+ {\it H.c.},
\label{ExtraV}
\end{eqnarray}
where $\Delta C(z, z^*)$ is a bilinear polynomial of $z$ and $z^*$.
The magnitude of third term and its hermitian conjugate
can be of order $m_{3/2}^3 M$, and so a large mixing mass of
Higgs doublets can be introduced.
Hence we need to impose the condition
\begin{eqnarray}
 \langle \tilde{F}^i \rangle
\frac{\partial \widehat{\cal W}}{\partial \tilde{z}^i}
= O(m_{3/2}^4)
\label{gh}
\end{eqnarray}
to guarantee the stability of weak scale.

The effective theories based on non-minimal SUGRA
with unified gauge symmetry also have been studied
in some literatures, but a complete analysis has not been
carried out yet.
For example, Hall {\it et} {\it al.} showed that the universality
of scalar masses is preserved
in the SUGRA whose K\"{a}hler potential has $U(n)$ symmetry
among the $n$ chiral fields\cite{HLW}.
Drees studied the low-energy theory based on SUGRA with
a non-canonical kinetic function
parametrized by one chiral field which triggers the SUSY
breaking\cite{Drees}.

As a recent development, the effective theory has been derived
from SUSY-GUT with a certain type of non-universal soft SUSY
breaking terms by integrating out superheavy fields\cite{KMY2}.
This starting SUSY-GUT can be derived from a certain type of
 non-minimal SUGRA with unified gauge symmetry by imposing
the {\it hidden} ansatz and taking the flat limit first.
The low-energy effective scalar potential is obtained as follows,
\begin{eqnarray}
V^{eff(non)} &=& V_{SUSY}^{eff(non)} + V_{Soft}^{eff(non)},
\label{Non-Veff}
\\
V_{SUSY}^{eff(non)} &=&
|\frac{\partial \widehat{\cal W}_{eff}}{\partial z^k}|^2
+ {1 \over 2}g_a^2 (z_k^* (T^a)^k_l z^l)^2,
\label{Non-VeffSUSY} \\
V_{Soft}^{eff(non)}&=& A \widehat{\cal W}_{eff}
+ B^k(z)_{eff} \frac{\partial \widehat{\cal W}_{eff}}
{\partial z^k}	+ {\it H.c.}
\nonumber \\
&~&	+ B^k(z)_{eff} {B_k(z)}_{eff} + C(z, z^*)_{eff}
 + \Delta V^{(non)},
\label{Non-VeffSoft}
\end{eqnarray}
where
$\widehat{\cal W}_{\it eff}$,
 $B^k(z)_{eff}$ and $C(z, z^*)_{eff}$ are
$\widehat{\cal W}$, $B^k(z)$ and $C(z, z^*)$
 with the extremum values for superheavy fields plugged in,
and $\Delta V^{(non)}$ is a sum of extra contributions.
There exist new contributions specific to SUSY-GUTs with
non-universal soft SUSY breaking terms.
The appearance of $D$-term contribution to the scalar masses
is one example.\footnote{
Historically, it was demonstrated that the $D$-term contribution
occurs when the gauge symmetry is broken at an intermediate scale
due to the non-universal soft scalar masses in
Refs.~\cite{Hagelin} and its existence in a more general situation
was suggested in Ref.~\cite{Faraggi}.}
In the absence of Fayet-Iliopoulos $D$-term, the conditions
that sizable $D$-term contributions appear are as follows.
(1) SUSY-GUT has non-universal soft SUSY breaking terms.
(2) The rank of the gauge group is reduced by the gauge symmetry
 breaking.
As the other feature, the gauge hierarchy achieved by a
fine-tuning in the superpotential would be violated, in general,
due to the non-universal SUSY breaking terms.
It is, however, shown that it is preserved for SUSY-GUT models
derived from the SUGRA with
the {\em hidden} ansatz and no light observable singlets.
It is also discussed some
phenomenological implications on the non-universal SUSY
breaking terms, including the utility of sfermion masses
as a probe of gauge symmetry breaking patterns and
the predictions of the radiative electroweak symmetry
breaking scenario and of no-scale type models.

As described in introduction,
it is important to study the low-energy theory in more general
framework of SUGRA because the SUSY spectrum can be a powerful
probe to the physics at higher energy scales.
The following subject has not been enough considered yet:
 to obtain the low-energy theory directly from non-minimal
SUGRA with unified gauge symmetry in model-independent manner.
In the following sections, we carry it out paying attention to
the gauge hierarchy problem.
And we discuss extra contributions to the SUSY breaking terms and
 the relation between our result and the previous one.

\section{Derivation of the Effective Lagrangian}

\subsection{Basic Assumptions}
\label{subsec:general-argument}

We have already explained general assumptions in the hidden
sector SUSY breaking scenario in subsection 2.1.
We shall first add basic assumptions although parts of them
would be repeated.
\begin{enumerate}
\item At the gravitational scale $M$,
the theory is described effectively as non-minimal
 SUGRA with a certain unified gauge symmetry
whose K\"ahler potential and superpotential are
given as
\begin{eqnarray}
K &=& {K}(z, z^*; \tilde{z}, \tilde{z}^*)
\nonumber \\
&=& \tilde{K}(\tilde{z}, \tilde{z}^*)
+ \Lambda(z, z^*; \tilde{z}, \tilde{z}^*)
\nonumber \\
&~& + {H}(z ; \tilde{z}, \tilde{z}^*) + H.c.
\label{K}
\end{eqnarray}
and
\begin{eqnarray}
W_{SG} &=& W_{SG}(z, \tilde{z})
\nonumber \\
&=& \tilde{W}(\tilde{z}) + W(z, \tilde{z}),
\label{W} \\
W(z, \tilde{z}) &\equiv& {1 \over 2}m_{\kappa\lambda}(\tilde{z})
z^{\kappa}z^{\lambda} + {1 \over 3!}f_{\kappa\lambda\mu}
(\tilde{z})z^{\kappa}z^{\lambda}z^{\mu} + \cdots ,
\end{eqnarray}
respectively.
Here the dots stands for terms of higher orders in $z$.
The gauge group is not necessarily grand-unified into a
simple group.
The theory has no Fayet-Iliopoulos $U(1)$ $D$-term
for simplicity.\footnote{
The extension of the theory with Fayet-Iliopoulos $D$-term
 is straightforward. We discuss it in Appendix B.}

\item The SUSY is spontaneously broken by the $F$-term
 condensation in the hidden sector.
The Planck scale physics plays an essential role
in the SUSY breaking.\footnote{
Our discussion is also applicable to the case of SUSY breaking
by gaugino condensation if the freedoms are effectively
replaced by some scalar multiplets whose VEVs are of order $M$
 as the models derived from SST.}
The hidden fields are gauge singlets
and they have the VEVs of $O(M)$.
The magnitude of ${W}_{SG}$ and $F$-component $\tilde{F}^i$ of
$\tilde{z}^i$ are $O(m_{3/2} M^2)$ and $O(m_{3/2} M)$, respectively.
%\footnote{A natural explanation that the VEV of $\tilde{W}$
%is $O(m_{3/2} M^2)$ has not been known yet.}

\item The unified gauge symmetry is broken at a scale $M_X$.
Some observable scalar fields have the VEVs of $O(M_X)$.

\item All the particles can be classified as heavy
(with mass $O(M_X)$) or light (with mass $O(m_{3/2})$).
The light observable fields are gauge non-singlets\footnote{
The reason why we assume it is that there is a difficulty that
radiative corrections generally induce a large tadpole contribution
 to Higgs masses in several models with a light singlet coupled to
Higgs doublets renormalizably in superpotential.}
 and have fluctuations only of $O(m_{3/2})$.
\end{enumerate}

\subsection{Vacuum Solutions}

The scalar potential is given as
\begin{eqnarray}
   V &=& V^{(F)} + V^{(D)},
\label{Vagain}
\\
V^{(F)} &\equiv& M^{2}exp(G/M^{2})(G^I (G^{-1})_I^J G_{J}-3M^{2})
\nonumber \\
 &\equiv& M^{2}exp(G/M^{2}) U,
\label{V(F)}
\\
V^{(D)} &\equiv& \frac{1}{2} (Re f^{-1})_{\alpha \beta}
D^{\alpha} D^{\beta}.
\label{V(D)}
\end{eqnarray}
The index $I$, $J$,... run all scalar species, $i$, $j$,... run
the hidden fields and $\kappa$, $\lambda$,... run the observable
 fields.
The $D^{\alpha}$'s are deformed as
\begin{eqnarray}
 D^\alpha &=&  K_\kappa (T^\alpha z)^\kappa
       = (z^{\dagger}T^\alpha)_\kappa K^\kappa
\label{Dagain}
\end{eqnarray}
from the gauge invariance of superpotential.\footnote{
Note that the superpotential is not gauge invariant
under Fayet-Iliopoulos $U(1)$ transformation.}
The vacuum $\langle z^I \rangle$
and $\langle z^*_J \rangle$ are determined
by solving the stationary conditions $\partial V /
\partial z^I = 0$ and $\partial V / \partial z^*_J = 0$.

The conditions that the SUSY is not spontaneously broken
in the observable sector are simply expressed as
\begin{eqnarray}
 \frac{\partial W}{\partial z^\kappa} &=& 0,
	\label{SUSYF}\\
{D}^\alpha &=& 0.
	\label{SUSYD}
\end{eqnarray}
We denote the solutions of the above conditions as
$z^{\kappa} = z_0^{\kappa}$.
We assume that our vacuum solution $\langle z^{\kappa} \rangle$ is
near to $z_0^{\kappa}$, i.e.,
$\langle z^{\kappa} \rangle = z_0^{\kappa}
+ O(m_{3/2})$.\footnote{
We can show that there exists at least such a vacuum solution
in the case that the scalar potential has no flat directions
in the SUSY limit.}

The supersymmetric fermion mass $\mu_{IJ}$ is given as
\begin{eqnarray}
\mu_{IJ} &=& \langle Me^{G/2M^2} ( G_{IJ} + {G_{I}G_{J} \over M^2}
 - G_{I'}(G^{-1})^{I'}_{J'} G^{J'}_{IJ} ) \rangle.
	\label{muIJ}
\end{eqnarray}
%and all the fields are classified by using $\mu_{IJ}$.
We take a basis of $z^I$ to diagonalize the SUSY
fermion mass matrix $\mu_{IJ}$.
Then we assume that the scalar fields are classified
either as ``heavy'' fields $z^K, z^L, \cdots$,  ``light'' fields
$z^k, z^l, \cdots$, $z^i, z^j, \cdots$ such as $\mu_{KL}=O(M_X)$,
$\mu_{kl}=O(m_{3/2})$, $\mu_{ij}=O(m_{3/2})$ or Nambu--Goldstone
fields $z^A, z^B, \cdots$ (which will be discussed just below).
It is shown that the hidden fields belong to the light sector
in Appendix A.

The mass matrix of the gauge bosons $(M_V^2)^{\alpha\beta}$ is
given as
\begin{eqnarray}
(M_V^2)^{\alpha\beta} =
  2 \langle (z^\dagger T^\beta)_\kappa K^\kappa_\lambda
  (T^\alpha z)^\lambda \rangle,
	\label{MV2}
\end{eqnarray}
up to the normalization due to the gauge coupling constants and
it can be diagonalized so that the gauge generators are
classified into ``heavy'' (those broken at $M_X$) $T^A, T^B,
\cdots$ and ``light'' (which remain unbroken above $m_{3/2}$)
$T^a, T^b, \cdots$. For  the heavy generators, the fields
$\langle (T^A z)^\kappa \rangle$ correspond to the directions of
the Nambu--Goldstone fields in the field space, which span a vector
space with the same dimension as the number of heavy generators.
We can take a basis of the Nambu--Goldstone multiplets, $z^A, z^B,
\cdots$ so that
\begin{eqnarray}
\sqrt{2} \langle (T^A z)^B \rangle = M_V^{AB} .
	\label{NG}
\end{eqnarray}
Here the Nambu--Goldstone fields are taken to be orthogonal to
the heavy and light fields such as $\langle (T^A z)^K
\rangle = 0$ and
$\langle (T^A z)^k \rangle = 0$.
To be more precise, either real or imaginary parts of
the $z^A$'s are the true Nambu--Goldstone bosons which are absorbed
into the gauge bosons, and the other parts acquire the same mass
 of order $M_X$ as that of the gauge bosons from the $D$-term
$V^{(D)}$ in the SUSY limit.
Hence the Nambu--Goldstone multiplets belong to the heavy sector.

Let us give the procedure to obtain the low-energy
effective theory.
\begin{enumerate}
\item  We calculate the VEVs of the derivatives of the potential
 and we write down the potential as
\begin{eqnarray}
 V=\frac{1}{2} \langle V_{IJ} \rangle
 \Delta z^I \Delta z^J +\cdots ,
\end{eqnarray}
where the scalar fields $z^I$'s are expanded
as $z^I = \langle z^I \rangle + \Delta z^I$
around the vacuum $\langle z^I \rangle$.

\item When there exists a mass mixing between the heavy and
light sectors, we need to diagonalize them to identify the
light and heavy fields correctly.

\item We solve the stationary conditions of the potential
for the heavy scalar fields while keeping the light scalar fields
arbitrary and then integrate out the heavy fields by inserting
the solutions of the stationary conditions into the potential.
We take the flat limit simultaneously.
\end{enumerate}

\subsection{Derivatives of $K$ and $W$}

It is convenient to
write both the K\"ahler potential $K$ and the superpotential
$W_{SG}$ in terms of the variations $\Delta z^I$ and
$\Delta z^*_J$ as follows,
\begin{eqnarray}
K &=& \langle K \rangle + \langle K_I \rangle \Delta z^I
+ \langle K^J \rangle
\Delta z^*_J
\nonumber \\
&~& + \langle K_I^J \rangle \Delta z^I \Delta z^*_J
\nonumber \\
&~& + {1 \over 2}\langle K_{IJ} \rangle \Delta z^I \Delta z^J
+ {1 \over 2}\langle K^{IJ} \rangle \Delta z^*_I \Delta z^*_J
\nonumber \\
&~&+ \cdots
	\label{expK}
\end{eqnarray}
and
\begin{eqnarray}
W_{SG} &=& \langle W_{SG} \rangle  + \langle W_{SG I} \rangle
 \Delta z^I
\nonumber \\
&~&+ {1 \over 2}\langle W_{SG IJ} \rangle \Delta z^I \Delta z^J
   + {1 \over 3!}\langle W_{SG IJJ'} \rangle \Delta z^I \Delta z^J
\Delta z^{J'}
\nonumber \\
&~&+ \cdots ,
	\label{expW}
\end{eqnarray}
where the ellipses represent higher order terms in $\Delta z$.

By using the expansions (\ref{expK}) and (\ref{expW}),
we find the following estimations
\begin{eqnarray}
&~& \langle G_i \rangle = O(M), \ \langle G_K \rangle \leq O(M_X),
\nonumber \\
&~& \langle G_A \rangle \leq O(m_{3/2}^2/M_X),
\ \langle G_k \rangle =0,
	\label{GI}\\
&~& \langle G_i^j \rangle \leq O(1), \
\langle G_{\kappa}^{\lambda} \rangle \leq O(1), \
\langle G_{K}^j \rangle \leq O(M_X/M),
\nonumber \\
&~& \langle G_{A}^j \rangle \leq O(M_X/M),
\ \langle G_{k}^j \rangle = 0,
	\label{GI^J}\\
&~& \langle G_{ij} \rangle \leq O(1), \
\langle G_{KL} \rangle \leq O(M_{KL}/m_{3/2}), \
\langle G_{Kj} \rangle \leq O(M_X/M),
\nonumber \\
&~& \langle G_{Aj} \rangle \leq O(M_X/M),
\ \langle G_{kj} \rangle = 0,
\nonumber \\
&~& \langle G_{\kappa B} \rangle \leq O(1),
\ \langle G_{\kappa l} \rangle
\leq O(1),
	\label{GIJ}
\end{eqnarray}
where $M_{KL}$ is the SUSY fermion mass coming
from the superpotential.
Here we used the assumption that our vacuum solution is near to
that in the SUSY limit and a perturbative argument
to derive the second relation in (\ref{GI}).
And we used the relations (\ref{NG}) and
$\langle D^{\alpha} \rangle
\leq O(m_{3/2}^2)$ to derive the third relation in (\ref{GI}).

By using the equality from the gauge invariance (\ref{G-inv2}),
we derive the following relations,
\begin{eqnarray}
&~&\langle G_{Akl} \rangle \leq O(1/M_X),
\ \langle G_{ABl} \rangle \leq O(1/M_X),
\nonumber \\
&~& \langle G_{ABC} \rangle \leq O(1/M_X),
\ \langle G_{Akj} \rangle \leq O(1/M),
\nonumber \\
&~& \langle G_{ABj} \rangle \leq O(1/M),
\ \langle G_{Aij} \rangle \leq O(1/M)
\label{GAIJ}
\end{eqnarray}
or
\begin{eqnarray}
&~&\langle {W_{SG}}_{Akl} \rangle \leq O(m_{3/2}/M_X),\
\langle {W_{SG}}_{ABl} \rangle \leq O(m_{3/2}/M_X),
\nonumber \\
&~& \langle {W_{SG}}_{ABC} \rangle \leq O(m_{3/2}/M_X),\
 \langle {W_{SG}}_{Akj} \rangle \leq O(m_{3/2}/M),
\nonumber \\
&~& \langle {W_{SG}}_{ABj} \rangle \leq O(m_{3/2}/M), \
 \langle {W_{SG}}_{Aij} \rangle \leq O(m_{3/2}/M).
\label{WAIJ}
\end{eqnarray}

\subsection{Stability of Gauge Hierarchy}

The mass squared matrices of the scalar fields are simply given
by the VEVs of the second derivatives of the potential. From
Eqs. (\ref{Vagain})--(\ref{Dagain}), we get the relations,
\begin{eqnarray}
 \lefteqn{V_{I}^{J}=
   \frac{\partial ^2 V}{\partial \phi^I \partial \phi^*_J}=}
\nonumber \\
  & & M^{2}(e^{G/M^{2}})_{I}^{J} U
     +M^{2}(e^{G/M^{2}})_I U^{J}
     +M^{2}(e^{G/M^{2}})^J U_I
     +M^{2}e^{G/M^{2}} U_{I}^{J}
\nonumber \\
   & &  +\frac{1}{2} (Re f^{-1})_{\alpha \beta, I}^{J}
 D^\alpha D^\beta +(Re f^{-1})_{\alpha \beta, I}
D^\alpha (D^\beta)^J +(Re f^{-1})_{\alpha \beta}^J
D^\alpha (D^\beta)_I
\nonumber \\
     & &     +(Re f^{-1})_{\alpha \beta }
 D^\alpha (D^\beta)_{I}^{J}+(Re f^{-1})_{\alpha \beta}
     (D^\alpha)_I (D^\beta)^J
\label{VI^J}
\end{eqnarray}
and
\begin{eqnarray}
 \lefteqn{V_{IJ}=
  \frac{\partial ^2 V}{\partial \phi^I \partial \phi^J}=}
\nonumber \\
  & & M^{2}(e^{G/M^{2}})_{IJ} U
     +M^{2}(e^{G/M^{2}})_I U_J
     +M^{2}(e^{G/M^{2}})_J U_I
     +M^{2}e^{G/M^{2}} U_{IJ}
\nonumber \\
   & &  +\frac{1}{2} (Re f^{-1})_{\alpha \beta, IJ}
       D^\alpha D^\beta+(Re f^{-1})_{\alpha \beta, I}
       D^\alpha (D^\beta)_J+(Re f^{-1})_{\alpha \beta, J}
            D^\alpha (D^\beta)_I
\nonumber \\
     & &     +(Re f^{-1})_{\alpha \beta }
 D^\alpha (D^\beta)_{IJ}+(Re f^{-1})_{\alpha \beta}
           (D^\alpha)_I (D^\beta)_J.
\label{VIJ}
\end{eqnarray}
By using the relations (\ref{GI})--(\ref{WAIJ}),
the VEVs of $V_{I}^{J}$ and $V_{IJ}$ are estimated as
\footnote{For simplicity, hereafter we consider only
the case that the equality holds.}
\begin{eqnarray}
&~&\langle {V^{(F)}}_{K}^L \rangle = O(M_X^2),
\ \langle {V^{(F)}}_{A}^B \rangle = O(m_{3/2}^2),\
\langle {V^{(F)}}_{k}^l \rangle = O(m_{3/2}^2),
\nonumber \\
&~&\langle {V^{(F)}}_{i}^j \rangle = O(m_{3/2}^2),\
\langle {V^{(F)}}_{K}^B \rangle = O(m_{3/2} M_X),\
\langle {V^{(F)}}_{K}^l \rangle = O(m_{3/2} M_X),
\nonumber \\
&~&\langle {V^{(F)}}_{K}^j \rangle = O(m_{3/2}M_X^2/M),\
\langle {V^{(F)}}_{A}^l \rangle = O(m_{3/2}^2),\
\nonumber \\
&~&\langle {V^{(F)}}_{A}^j \rangle = O(m_{3/2}^2 M_X/M),\
\langle {V^{(F)}}_{k}^j \rangle = 0
\label{<VI^J>F}
\end{eqnarray}
and
\begin{eqnarray}
&~&\langle V^{(F)}_{KL} \rangle = O(m_{3/2} M),
\ \langle V^{(F)}_{AB} \rangle = O(m_{3/2}^2),\
\langle V^{(F)}_{kl} \rangle = O(m_{3/2} M),
\nonumber \\
&~&\langle V^{(F)}_{ij} \rangle = O(m_{3/2} M),\
\langle V^{(F)}_{KB} \rangle = O(m_{3/2} M),\
\langle V^{(F)}_{Kl} \rangle = O(m_{3/2} M),
\nonumber \\
&~&\langle V^{(F)}_{Kj} \rangle = O(m_{3/2} M),\
\langle V^{(F)}_{Al} \rangle = O(m_{3/2}^2),\
\langle V^{(F)}_{Aj} \rangle = O(m_{3/2}^2),
\nonumber \\
&~&\langle V^{(F)}_{kj} \rangle = 0,
\label{<VIJ>F}
\end{eqnarray}
respectively.
The quantities of order $m_{3/2} M$ in
$\langle V^{(F)}_{IJ} \rangle$
 originates in the term $\langle Me^{G/2M^{2}}G_{IJJ'} \rangle
\langle F^{J'} \rangle$.
If $\langle V_{IJ} \rangle$'s are $O(m_{3/2} M)$ for the light
fields $z^I$, the masses of light fields can get intermediate
values after the diagonalization of mass matrix.
The masses of those fermionic partners stay at the weak scale.
The weak scale can be destabilized in the presence of the weak
Higgs doublets with intermediate masses.
This is so called $\lq$gauge hierarchy problem'.
Only when $\langle Me^{G/2M^{2}}G_{IJJ'} \rangle
 \langle F^{J'} \rangle$'s meet
some requirements, the hierarchy survives.
In this paper, we require the following conditions,
\begin{eqnarray}
\langle Me^{G/2M^{2}}G_{IJJ'} \rangle \langle {F}^{J'} \rangle \leq
O(m_{3/2}^2)
\label{gh1}
\end{eqnarray}
for the light fields $z^I$ and $z^J$,
\begin{eqnarray}
&~&    \langle Me^{G/2M^{2}}G_{KlJ'} \rangle
 \langle {F}^{J'} \rangle \leq
O(m_{3/2} M_X),
\label{gh3} \\
&~&    \langle Me^{G/2M^{2}}G_{KjJ'} \rangle
 \langle {F}^{J'} \rangle \leq
O(m_{3/2} M_X^2/M),
\label{gh4} \\
&~&    \langle Me^{G/2M^{2}}G_{IJJ'} \rangle
\langle {F}^{J'} \rangle \leq
O(M_X^2)
\label{gh2}
\end{eqnarray}
for the heavy fields $z^I$ and $z^J$, and
\begin{eqnarray}
    \langle Me^{G/2M^{2}}G_{AjJ'} \rangle
\langle {F}^{J'} \rangle \leq
O(m_{3/2}^2 M_X/M).
\label{gh5}
\end{eqnarray}
The conditions (\ref{gh1})--(\ref{gh5}) correspond to
the statement that the magnitudes of
$\langle Me^{G/2M^{2}}G_{IJJ'} \rangle \langle F^{J'} \rangle$
 are equal to or smaller than the rest terms.
The {\it hidden} ansatz trivially satisfies the above conditions.
The gauge hierarchy problem has been discussed on the postulation
that $M_X$ is identified with $M$ in Ref.~\cite{JKY}.

The contributions from the $D$-term are naively estimated
as follows,
\begin{eqnarray}
&~&\langle {V^{(D)}}_{\kappa}^{\lambda} \rangle = O(M_X^2),
\ \langle V^{(D)}_{\kappa\lambda} \rangle = O(M_X^2),
\nonumber \\
&~&\langle {V^{(D)}}_{K}^j \rangle,
\ \langle {V^{(D)}}_{A}^j \rangle = O(M_X^3/M),
\ \langle V^{(D)}_{Kj} \rangle,
\ \langle V^{(D)}_{Aj} \rangle = O(M_X^3/M),
\nonumber \\
&~&\langle {V^{(D)}}_{i}^j \rangle,
\ \langle V^{(D)}_{ij} \rangle = O(M_X^4/M^2),
\nonumber \\
&~&\langle {V^{(D)}}_{k}^j \rangle,
\ \langle V^{(D)}_{kj} \rangle = 0,
\label{<VIJ>D}
\end{eqnarray}
where we used the relation $\langle (D^A)_I \rangle
= \langle (z^{\dagger} T^A)_{\kappa} K_I^{\kappa} \rangle
= O(M_X) K^A_I$.
We require that the light fields defined by using $\mu_{IJ}$
 get no heavy masses from the $D$-term.
For simplicity, we impose the conditions such as
\begin{eqnarray}
\langle {V^{(D)}}_I^J \rangle \leq \langle {V^{(F)}}_I^J \rangle
\label{<VD><<VF>1}
\end{eqnarray}
and
\begin{eqnarray}
\langle V^{(D)}_{IJ} \rangle \leq \langle V^{(F)}_{IJ} \rangle
\label{<VD><<VF>2}
\end{eqnarray}
for the light fields $z^I$.
They yield to the following relations
\begin{eqnarray}
\langle K_{A}^{k} \rangle, \ \langle K_{Ak} \rangle
= O({m_{3/2}^2 \over M_X^2})
\label{<KAk>}
\end{eqnarray}
and
\begin{eqnarray}
\langle K_{A}^{i} \rangle, \ \langle K_{Ai} \rangle
= O({m_{3/2}^2 \over M_XM}).
\label{<KAi>}
\end{eqnarray}
The analysis could be made based on weaker requirements than
 (\ref{gh1})--(\ref{gh5}), (\ref{<VD><<VF>1}) and
(\ref{<VD><<VF>2}), but we will not discuss it further to avoid
a complication and a subtlety in this paper.

\subsection{Diagonalization of Mass Matrix}

The mass term is written as
\begin{eqnarray}
 V^{mass}=\frac{1}{2} \langle V_{\hat{I}\hat{J}} \rangle
{\Delta z}^{\hat{I}}{\Delta z}^{\hat{J}}  ,
\label{Vmass}
\end{eqnarray}
where ${\Delta z}^{\hat{I}}$$=({\Delta z}^K, {\Delta z}^{\bar{L}}
\equiv {\Delta z}^*_L;
 {\Delta z}^A, {\Delta z}^{\bar{B}} \equiv  {\Delta z}^*_B;$
$\Delta \tilde{z}^i,$ $\Delta \tilde{z}^{\bar{j}}$ $\equiv$
$\Delta \tilde{z}^*_j;$ ${\Delta z}^k,$ ${\Delta z}^{\bar{l}}$
$\equiv$ ${\Delta z}^*_l)$. From the discussion in the previous
subsection, the orders of $\langle V_{\hat{I}\hat{J}} \rangle$
are estimated as
\begin{eqnarray}
 \langle V_{\hat{I}\hat{J}} \rangle = O
\left(
\begin{array}{ccc}
M_X^2 & M_X^2 & m_{3/2} M_X \\
M_X^2 & M_X^2 & m_{3/2}^2  \\
m_{3/2} M_X & m_{3/2}^2 & m_{3/2}^2
\end{array}
\right)
\label{<VIJ>N}
\end{eqnarray}
for gauge non-singlet fields $({\Delta z}^{\hat{K}};
 {\Delta z}^{\hat{A}}; {\Delta z}^{\hat{k}})$
 and
\begin{eqnarray}
 \langle V_{\hat{I}\hat{J}} \rangle = O
\left(
\begin{array}{ccc}
M_X^2 & M_X^2 & m_{3/2} M_X^2/M \\
M_X^2 & M_X^2 & m_{3/2}^2 M_X/M \\
m_{3/2} M_X^2/M & m_{3/2}^2 M_X/M & m_{3/2}^2
\end{array}
\right)
\label{<VIJ>S}
\end{eqnarray}
for gauge singlet fields $({\Delta z}^{\hat{K}};
{\Delta z}^{\hat{A}}; \Delta \tilde{z}^{\hat{i}})$.
As the matrix $\langle V_{\hat{I}\hat{J}} \rangle$ is hermitian,
it can be diagonalized by the use of a certain unitary matrix
$U^{\hat{I}}_{\hat{J}}$.
The mass eigenstate $\phi^{\hat{I}}$ is related to
${\Delta z}^{\hat{I}}$ as $\phi^{\hat{I}} = U^{\hat{I}}_{\hat{J}}
{\Delta z}^{\hat{J}}$.
We denote the heavy fields with mass $O(M_X)$ as
$\phi^{\hat{\cal H}}$ and the light fields with mass $O(m_{3/2})$
as $\phi^{\hat{\cal L}}$ where ${\cal H} = (K, A)$ and ${\cal L}
= (i, k)$. Next we would like to integrate out the heavy fields
$\phi^{\hat{\cal H}}$.
For this purpose, it is convenient to choose the variables
\begin{eqnarray}
\Delta \hat{z}^{\hat{\cal H}} &=&
(U^{\hat{\cal H}}_{\hat{\cal H'}})^{-1} \phi^{\hat{\cal H'}},
\label{hatzH}
\\
\Delta \hat{z}^{\hat{\cal L}} &=&
(U^{\hat{\cal L}}_{\hat{\cal L'}})^{-1} \phi^{\hat{\cal L'}}
\label{hatzL}
\end{eqnarray}
or
\begin{eqnarray}
\Delta \hat{z}^{\hat{I}} &=& \hat{U}^{\hat{I}}_{\hat{J}}
 {\Delta z}^{\hat{J}},
\label{hatz}
\\
\hat{U}^{\hat{I}}_{\hat{J}} &\equiv&
\left(
\begin{array}{cc}
I & (U^{\hat{\cal H}}_{\hat{\cal H'}})^{-1}
U^{\hat{\cal H'}}_{\hat{\cal L}}\\
(U^{\hat{\cal L}}_{\hat{\cal L'}})^{-1}
U^{\hat{\cal L'}}_{\hat{\cal H}}
 & I
\end{array}
\right).
\label{hatU}
\end{eqnarray}
Here we used the fact that $det U^{\hat{\cal H}}_{\hat{\cal H'}} =
1 + O(m_{3/2}^2/M_X^2)$ and $det U^{\hat{\cal L}}_{\hat{\cal L'}} =
1 + O(m_{3/2}^2/M_X^2)$ and neglected the higher order terms.
% of $O(m_{3/2}^2/M_X^2)$.
The orders of off-diagonal elements of
$\hat{U}^{\hat{I}}_{\hat{J}}$ are estimated as
\begin{eqnarray}
&~&\hat{U}^{\hat{K}}_{\hat{l}}
    = O({m_{3/2} \over M_X}),~
\hat{U}^{\hat{A}}_{\hat{l}}
    = O({m_{3/2}^2 \over M_X^2}),
\label{off1}
\\
&~&\hat{U}^{\hat{K}}_{\hat{j}}
    = O({m_{3/2} \over M}),~
\hat{U}^{\hat{A}}_{\hat{j}}
    = O({m_{3/2}^2 \over M M_X}).
\label{off2}
\end{eqnarray}

\subsection{Calculation of the Effective Theory}

The rest in the procedure are as follows,\\
1. We write down the scalar potential by using new variables
 $\Delta \hat{z}^{\hat{I}}$.\\
2. We take the flat limit and integrate out the heavy fields
by inserting the solutions of the stationary conditions into
the full potential.

We can write down the K\"ahler potential $K$, the superpotential
$W_{SG}$ and the $D$-auxiliary fields $D^\alpha$ in terms of
the variations $\Delta \hat{z}^{\hat{I}}$ as follows,
\begin{eqnarray}
K &=&
\hat{K}(\Delta \hat{z})
\nonumber \\
&=&\langle \hat{K} \rangle + \langle \hat{K}_{\hat{I}} \rangle
 \Delta \hat{z}^{\hat{I}}  + {1 \over 2}\langle
\hat{K}_{\hat{I}\hat{J}} \rangle
\Delta \hat{z}^{\hat{I}} \Delta \hat{z}^{\hat{J}}
+ \cdots ,
	\label{hatK}\\
%\end{eqnarray}
%and
%\begin{eqnarray}
W_{SG} &=&
\hat{W}(\Delta \hat{z})
\nonumber \\
&=&\langle \hat{W} \rangle + \langle \hat{W}_{\hat{I}} \rangle
 \Delta \hat{z}^{\hat{I}}  + {1 \over 2}\langle
\hat{W}_{\hat{I}\hat{J}} \rangle
\Delta \hat{z}^{\hat{I}} \Delta \hat{z}^{\hat{J}}
\nonumber \\
&~&   + {1 \over 3!}\langle \hat{W}_{\hat{I}\hat{J}\hat{J'}}
\rangle \Delta \hat{z}^{\hat{I}} \Delta \hat{z}^{\hat{J}}
\Delta \hat{z}^{\hat{J'}}
+ \cdots
	\label{hatW}
\end{eqnarray}
and
\begin{eqnarray}
D^{\alpha} &=& \hat{D}^{\alpha}(\Delta \hat{z})
\nonumber \\
&=& (\hat{K}_\lambda  +  \hat{K}_{\hat{I}}
\Delta \hat{U}^{\hat{I}}_{\lambda}) (T^A)_\kappa^\lambda
(\langle z^{\kappa} \rangle +
(\hat{U}^{-1})^{\kappa}_{\hat{J}}\Delta \hat{z}^{\hat{J}}) ,
\label{hatD}
\end{eqnarray}
where the ellipses represent terms of higher orders and
 $\hat{U}^{\hat{I}}_{\hat{J}} =
\delta^{\hat{I}}_{\hat{J}} + \Delta \hat{U}^{\hat{I}}_{\hat{J}}$.

For a later convenience, we deform $V^{(F)}$ as follows,
\begin{eqnarray}
   V^{(F)} &=&
exp(\hat{K}/M^{2})\biggl(\widehat{{\cal G}}_{\bar{\kappa}}
(\hat{K}^{-1})^{{\bar{\kappa}}{\lambda}}
 \widehat{{\cal G}}_{\lambda}
\nonumber \\
&~&~~~~~~ + \hat{\cal G}_{\bar{i}} (\widehat{K}^{-1})^{{\bar{i}}j}
 \hat{\cal G}_{j}
 -3{|\hat{W}|^2 \over M^{2}}\biggr) + \Delta V^{(F)},
\label{V(F)again}
\end{eqnarray}
where
\begin{eqnarray}
\widehat{{\cal G}}_{\bar{\kappa}} &\equiv&
\hat{\cal G}_{\bar{\kappa}} + \hat{\cal G}_{\bar{i}}
(\hat{K}^{-1})^{{\bar{i}}\mu} (\hat{K})_{{\mu}{\bar{\kappa}}},
\label{widehatg*}
\\
\widehat{\cal G}_{\lambda} &\equiv&
\hat{\cal G}_{\lambda} + (\hat{K})_{\lambda\bar{\nu}}
 (\hat{K}^{-1})^{{\bar{\nu}}j} \hat{\cal G}_j,
\label{widehatg}
\\
\hat{\cal G}_{\bar{I}} &\equiv& \hat{W}^*_{\bar{I}}
+ {\hat{K}_{\bar{I}} \over M^2}\hat{W}^* ,
\label{hatg*}
\\
\hat{\cal G}_{I} &\equiv& \hat{W}_{I}
+ {\hat{K}_{I} \over M^2}\hat{W}
\label{hatg}
\end{eqnarray}
and
\begin{eqnarray}
(\widehat{K}^{-1})^{{\bar{i}}j} &\equiv&
(\hat{K}^{-1})^{{\bar{i}}j}
- (\hat{K}^{-1})^{{\bar{i}}\mu} (\hat{K})_{\mu\bar{\nu}}
 (\hat{K}^{-1})^{{\bar{\nu}}j},
\label{widehatK}
\\
\Delta V^{(F)} &\equiv&
exp(\hat{K}/M^{2})\biggl(\hat{\cal G}_{\hat{I}}
\Delta(\hat{U})^{\hat{I}}_{\bar{I}}
(\hat{K}^{-1})^{{\bar{I}}J} \hat{\cal G}_J
+\hat{\cal G}_{\bar{I}}(\hat{K}^{-1})^{{\bar{I}}J}
\hat{\cal G}_{\hat{J}} \Delta(\hat{U})^{\hat{J}}_J
\nonumber \\
&~&~~~~ +\hat{\cal G}_{\bar{I}}[(\hat{K}^{-1})^{{\hat{J}}J}
\Delta(\hat{U}^{-1})_{\hat{J}}^{\bar{I}}
+ (\hat{K}^{-1})^{{\hat{I}}{\bar{I}}}
\Delta(\hat{U}^{-1})^J_{\hat{I}}] \hat{\cal G}_{J} \biggr)
\nonumber \\
&~&~~~~ + O((\Delta U^{(-1)})^2).
\label{DeltaV(F)}
\end{eqnarray}
We should not confuse $(\widehat{K}^{-1})^{{\bar{i}}j}$,
$\widehat{{\cal G}}_{\bar{\kappa}}$
and $\widehat{\cal G}_{\lambda}$ with
$(\hat{K}^{-1})^{{\bar{i}}j}$, $\hat{{\cal G}}_{\bar{\kappa}}$
and $\hat{\cal G}_{\lambda}$, respectively.
(Notice that the difference of the size of hat.)
Here $(\hat{K})_{\mu{\bar{\nu}}}$ is
the inverse matrix of $(\hat{K}^{-1})^{\mu{\bar{\nu}}}$.
We expand $\Delta \hat{z}^{\hat{I}}$ in powers of $m_{3/2}$ such as
\begin{eqnarray}
\Delta \hat{z}^{\hat{I}} &=&
\delta \hat{z}^{\hat{I}} + \delta^2 \hat{z}^{\hat{I}}
+ \cdots,
\label{exphatz}
\end{eqnarray}
with $\delta^n \hat{z}^{\hat{I}} = O(m_{3/2}^n / M_X^{n-1})$.
We assume $\Delta \hat{z}^{\hat{k}} =
O(m_{3/2})$ for the light fields, {\it e.g.},\/
$\delta^2 \hat{z}^{\hat{k}} = \delta^3 \hat{z}^{\hat{k}}
= \cdots =0$.
In the same way, we expand the $\widehat{\cal G}_{\hat{\lambda}}$,
 $\hat{\cal G}_{\hat{j}}$ and $\hat{D}^\alpha$
 in powers of $m_{3/2}$ such as
\begin{eqnarray}
\widehat{\cal G}_{\hat{\lambda}} &=&
 \delta \widehat{\cal G}_{\hat{\lambda}}
+ \delta^2 \widehat{\cal G}_{\hat{\lambda}} + \cdots,
 	\label{expwidehatg}\\
\hat{\cal G}_{\hat{j}} &=&
 \delta \hat{\cal G}_{\hat{j}} + \delta^2 \hat{\cal G}_{\hat{j}}
+ \cdots 	\label{exphatg}
\end{eqnarray}
and
\begin{eqnarray}
\hat{D}^\alpha &=& \delta
\hat{D}^\alpha + \delta^2 \hat{D}^\alpha + \cdots .
\label{exphatD}
\end{eqnarray}
Those orders are given as $\delta^n \widehat{\cal G}_{\hat{\lambda}}
= O(m_{3/2}^n / M_X^{n-2})$,
$\delta^n \hat{\cal G}_{\hat{j}} = O(m_{3/2}^n / M^{n-2})$
and  $\delta^n \hat{D}^\alpha$$=$$O(m_{3/2}^n / M_X^{n-2})$
up to the factor $O((M_X/M)^n)$.

The following relations are derived from the expansions of
$\widehat{\cal G}_\lambda$ and $\hat{\cal G}_{j}$
\begin{eqnarray}
\delta \widehat{\cal G}_K &=& \langle \hat{W}_K \rangle +
 \langle \hat{W}_{KL} \rangle \delta \hat{z}^L
\nonumber \\
&~&+ {\langle \hat{W} \rangle \over M^2}\langle \hat{K}_K \rangle
+ \langle (\hat{K})_{K\bar{\nu}} \rangle \langle
(\hat{K}^{-1})^{{\bar{\nu}}j} \rangle \delta \hat{\cal G}_j,
\label{gK1}
\\
\delta^2 \widehat{\cal G}_K &=&
\langle \hat{W}_{KL} \rangle \delta^2 \hat{z}^L
+ \langle \hat{W}_{K{\cal L}} \rangle \delta \hat{z}^{\cal L}
+ \langle \hat{W}_{KA} \rangle \delta \hat{z}^A
\nonumber \\
&~& + {1 \over 2}\langle \hat{W}_{K\lambda\mu} \rangle
\delta \hat{z}^{\lambda} \delta \hat{z}^{\mu}
\nonumber \\
&~& + {1 \over M^2}\biggl({1 \over 2}\langle \hat{W}_{LM} \rangle
\delta \hat{z}^L \delta \hat{z}^M \langle \hat{K}_K \rangle
+ \langle \hat{W} \rangle \langle \hat{K}_{K\hat{J}} \rangle
\delta \hat{z}^{\hat{J}}\biggr)
 \nonumber \\
&~& + \langle (\hat{K})_{K\bar{\nu}} \rangle
\langle (\hat{K}^{-1})^{{\bar{\nu}}j} \rangle
\delta^2 \hat{\cal G}_j
 + \delta ((\hat{K})_{K\bar{\nu}} (\hat{K}^{-1})^{{\bar{\nu}}j})
 \delta \hat{\cal G}_j,
\label{gK2}
\\
\delta \widehat{\cal G}_A &=&  \langle (\hat{K})_{A\bar{\nu}}
\rangle \langle (\hat{K}^{-1})^{{\bar{\nu}}j} \rangle
\delta \hat{\cal G}_j,
\label{gA1}
\\
\delta^2 \widehat{\cal G}_A &=&
%    \langle \hat{W}_{A} \rangle +
 \langle \hat{W}_{AI} \rangle \delta \hat{z}^{I}
%\nonumber \\
 + {1 \over 2}\langle \hat{W}_{AKI} \rangle
\delta \hat{z}^{K} \delta \hat{z}^{I}
+ {\langle \hat{W} \rangle \over M^2}
\langle \hat{K}_{A\hat{J}} \rangle \delta \hat{z}^{\hat{J}}
 \nonumber \\
&~& + \langle (\hat{K})_{A\bar{\nu}} \rangle
\langle (\hat{K}^{-1})^{{\bar{\nu}}j} \rangle
\delta^2 \hat{\cal G}_j
 + \delta ((\hat{K})_{A\bar{\nu}} (\hat{K}^{-1})^{{\bar{\nu}}j})
 \delta \hat{\cal G}_j
\label{gA2}
%\\
%   \delta^3 {\cal{cal G}_A} &=&
% h_{A\lambda\mu} \delta z^\lambda \delta^2 z^\mu
%+ {\tilde{W} \over M^2}\delta^2 K_{A}
% \nonumber \\
%&~& + (K)_{A0}^\nu (K^{-1})_{\nu 0}^j \delta^3 \hat{cal G}_j
% + \delta ((K)_{A}^\nu (K^{-1})_{\nu}^j) \delta^2 \hat{cal G}_j
% \nonumber \\
%&~& + \delta^2 ((K)_{A}^\nu (K^{-1})_{\nu}^j) \delta \hat{cal G}_j.
%\label{calGA3rd}
\end{eqnarray}
and
\begin{eqnarray}
   \delta {\hat{\cal G}_j} &=&  \langle \hat{W}_j \rangle
+{\langle \hat{W} \rangle \over M^2} \langle \hat{K}_{j} \rangle,
\label{gj1}
\\
   \delta^2 {\hat{\cal G}_j} &=&
 \langle \hat{W}_{jI} \rangle \delta \hat{z}^{I}
 + {1 \over 2}\langle \hat{W}_{jIJ} \rangle
\delta \hat{z}^{I} \delta \hat{z}^{J}
\nonumber \\
&~& + {1 \over M^2}\biggl({1 \over 2}\langle \hat{W}_{LM} \rangle
\delta \hat{z}^L \delta \hat{z}^M \langle \hat{K}_j \rangle
+ \langle \hat{W} \rangle
\langle \hat{K}_{j\hat{J}} \rangle
\delta \hat{z}^{\hat{J}}\biggr),
\label{gj2}
% \\
%   \delta^3 {\hat{\cal G}_j} &=&
% {1 \over 2}\partial_j \mu_{kl} \delta z^k \delta z^l
% \partial_j \mu_{\kappa\lambda}
% \delta z^\kappa \delta^2 z^\lambda
% + {1 \over 3!}\partial_j h_{\lambda\mu\nu} \delta z^\lambda
% \delta z^\mu \delta z^\nu
% \nonumber \\
%&~& +{1 \over M^2}
% ({1 \over 2}\mu_{kl} \delta z^k \delta z^l
% +  \mu_{LM} \delta z^L \delta^2 z^M
% + {1 \over 3!} h_{\lambda\mu\nu} \delta z^\lambda
% \delta z^\mu \delta z^\nu) K_j
% \nonumber \\
%&~& + {\tilde{W} \over M^2}\delta^2 K_{j} ,
%\label{hatGj3rd}
\end{eqnarray}
respectively.
While the expansion of $\hat{D}^A$ gives
\begin{eqnarray}
\delta \hat{D}^A &=&
\langle \hat{K}_\lambda \rangle (T^A)_\kappa^\lambda
\delta \hat{z}^\kappa
 \nonumber \\
&~&+ (\langle \hat{K}_{{\hat{I}}\lambda} \rangle \delta
\hat{z}^{\hat{I}} + \langle \hat{K}_{\hat{I}} \rangle \delta
\hat{U}_{\lambda}^{\hat{I}})
 (T^A)_\kappa^\lambda \langle z^\kappa \rangle,
\label{D1}\\
\delta^2 \hat{D}^A &=&
\langle \hat{K}_\lambda \rangle (T^A)_\kappa^\lambda
(\delta^2 \hat{z}^\kappa + \delta(\hat{U}^{-1})_{\hat{J}}^\kappa
\delta \hat{z}^{\hat{J}})
 \nonumber \\
&~&+ ( \langle \hat{K}_{{\hat{I}}\lambda} \rangle \delta^2
\hat{z}^{\hat{I}} + \langle \hat{K}_{{\hat{I}}{\hat{J}}\lambda}
\rangle \delta \hat{z}^{\hat{I}} \delta \hat{z}^{\hat{J}}
\nonumber \\
&~&~~~~ + \langle \hat{K}_{\hat{I}} \rangle \delta^2
\hat{U}_{\lambda}^{\hat{I}}) (T^A)_\kappa^\lambda
\langle z^\kappa \rangle
\nonumber \\
&~&+ ( \langle \hat{K}_{{\hat{I}}\lambda} \rangle
\delta \hat{z}^{\hat{I}}
 +\langle \hat{K}_{\hat{I}} \rangle \delta
\hat{U}_{\lambda}^{\hat{I}})
 (T^A)_\kappa^\lambda \delta \hat{z}^\kappa .
\label{D2}
\end{eqnarray}
The expansions of the stationary conditions
$\partial V/\partial z^K=0$ and
$\partial V/\partial z^A=0$ give
\begin{eqnarray}
  \langle \hat{W} \rangle_{KL} \langle
(\hat{K}^{-1})^{L{\bar{\mu}}} \rangle
 \delta \widehat{{\cal G}}_{\bar{\mu}} &=& 0,
\label{stK1}\\
  \langle \hat{W} \rangle_{KL} \langle
(\hat{K}^{-1})^{L{\bar{\mu}}} \rangle
 \delta^2 \widehat{{\cal G}}_{\bar{\mu}}
 &=&
 - \delta \widehat{{\cal G}}_{\bar{\mu}} \langle
(\hat{K}^{-1})^{{\bar{\mu}}\lambda} \rangle
 \langle \hat{W}_{\lambda\sigma K} \rangle \delta \hat{z}^{\sigma}
%+ (K)_{\lambda 0}^\nu (K^{-1})_{\nu 0}^j)
%\partial_j \mu_{K \lambda} \delta z^\lambda\}
+ const.
\label{stK2}
\end{eqnarray}
and
\begin{eqnarray}
\langle Re f_{\alpha\beta}^{-1} \rangle \langle
(\hat{z} T^\alpha)^{\bar{\mu}} \rangle \langle
\hat{K}_{A\bar{\mu}} \rangle \delta \hat{D}^\beta &=& 0,
\label{stA1}\\
\langle Re f_{\alpha\beta}^{-1} \rangle \langle
(\hat{z} T^\alpha)^{\bar{\mu}} \rangle \langle
\hat{K}_{A\bar{\mu}} \rangle \delta^2 D^\beta &=&
 E \delta \widehat{{\cal G}}_{\bar{\mu}}
\langle (K^{-1})^{{\bar{\mu}}\lambda} \rangle
\langle \hat{W}_{\lambda \sigma A} \rangle
\delta \hat{z}^\sigma \\
\nonumber
&~&+ const.,
\label{stA2}
\end{eqnarray}
respectively.
Here $E \equiv \langle exp(K/M^2) \rangle$.

{}From Eqs.~(\ref{gK1}), (\ref{gA1}),
(\ref{gj1}) and (\ref{stK1}), we find $\delta \hat{z}^K =0$ by
using $\langle \delta \hat{z}^K \rangle =0$.
Eq.~(\ref{stK2}) gives the solution for $\delta^2
\widehat{{\cal G}}_{\bar{K}}$ as
\begin{eqnarray}
   \delta^2 \widehat{{\cal G}}_{\bar{K}} &=&
\langle \widehat{{\cal G}}_{\bar{K}} \rangle
-\langle (\hat{K}_{{\bar{K}}L}) \rangle
\langle \hat{W}^{-1} \rangle^{KL}
\delta \widehat{{\cal G}}_{\bar{\mu}} \langle
(K^{-1})^{{\bar{\mu}}M} \rangle  \langle \hat{W}_{MKl}
\rangle \delta \hat{z}^l
\label{<gK>}
\end{eqnarray}
where a constant factor of $\delta^2 \widehat{{\cal G}}_{\bar{K}}$
is denoted as $\langle \widehat{{\cal G}}_{\bar{K}} \rangle$. From
Eqs.~(\ref{off2}), (\ref{D1}) and (\ref{stA1}),
 we find $\delta \hat{D}^A = 0$ and $\delta \hat{z}^A = 0$.
By using the relations $\langle \hat{W}_{ABk} \rangle
=O(m_{3/2}/M_X)$ and
$\langle \hat{W}_{ABi} \rangle=O(m_{3/2}/M)$,
%as a consequence of the gauge invariance,
we can show that $\delta^2 \hat{D}^A$ is a
constant independent of the light fields.
Therefore we will denote it by $\langle \hat{D}^A \rangle$.

Now it is straightforward to calculate the scalar potential
 ${\cal V}^{eff}$ in the low-energy effective theory by
substituting the solutions of the stationary conditions
for the heavy fields.
The result can be compactly expressed if we define the effective
superpotential $\widehat{\cal W}_{eff}$ as
\begin{eqnarray}
    \widehat{\cal W}_{eff} (z) &=&
  {1 \over 2!}\hat{\mu}_{kl} \delta \hat{z}^k \delta \hat{z}^l
+  {1 \over 3!}\hat{h}_{klm} \delta \hat{z}^k \delta \hat{z}^l
\delta \hat{z}^m ,
\label{calWeff}
\end{eqnarray}
where
\begin{eqnarray}
  \hat{\mu}_{kl} &\equiv& E^{1/2}\biggl(\langle
\hat{W}_{kl} \rangle + {\langle \hat{W} \rangle \over M^2}
\langle \hat{K}_{kl} \rangle
 - \langle \hat{K}_{kl\bar{i}} \rangle \langle
(\hat{K}^{-1})^{{\bar{i}}j} \rangle
 \delta \hat{\cal G}_j \biggr)
\nonumber \\
&~&~~~~ + (m^{'''}_{3/2})_{kl} ,
\label{hat-mu}\\
 \hat{h}_{klm} &\equiv& E^{1/2} \langle \hat{W}_{klm} \rangle.
\label{hat-h}
\end{eqnarray}
Then we can write down the scalar potential of effective
theory as\footnote{
Here we omitted the terms irrelevant to the gauge non-singlet
fields $\delta \hat{z}^{\hat{k}}$ and the terms whose magnitudes
are less than $O(m_{3/2}^4)$.}
\begin{eqnarray}
{\cal V}^{eff}&=& {\cal V}_{SUSY}^{eff} + {\cal V}_{Soft}^{eff},
\label{calVeff}
\\
{\cal V}_{SUSY}^{eff} &=&
\frac{\partial \widehat{\cal W}^*_{eff}}{\partial \hat{z}^{\bar{k}}}
 \langle (\hat{K}^{-1})^{{\bar{k}}l} \rangle
\frac{\partial \widehat{\cal W}_{eff}}{\partial \hat{z}^l}
+ {1 \over 2}g_a^2 (\langle \hat{K}_{k\bar{l}} \rangle
\hat{z}^{\bar{l}} (T^a)^k_l \hat{z}^l)^2,
\label{calVeffSUSY}
\\
{\cal V}_{Soft}^{eff}&=& A \widehat{\cal W}_{eff}
   + B_{\bar{k}}(\hat{z})_{eff}
 \langle (\hat{K}^{-1})^{{\bar{k}}l} \rangle
\frac{\partial \widehat{\cal W}_{eff}}{\partial \hat{z}^l}
		+ {\it H.c.}
\nonumber \\
&~&+ B_{\bar{k}}(\hat{z})_{eff}
 \langle (\hat{K}^{-1})^{{\bar{k}}l} \rangle
{B_l(\hat{z})}_{eff} + C(\hat{z})_{eff}
\nonumber \\
&~& + \Delta \widehat{\cal V} + \Delta {\cal V}'^{(F)} ,
\label{calVeffsoft}
\end{eqnarray}
where $\Delta \widehat{\cal V} + \Delta {\cal V}'^{(F)}$
 is a sum of contributions such as
\begin{eqnarray}
\Delta \widehat{\cal V} &=& \Delta \widehat{\cal V}_0^{(F)}
 + \Delta \widehat{\cal V}_0^{(D)} + \Delta
\widehat{\cal V}_1^{(F)} + \Delta \widehat{\cal V}_1^{(D)} ,
\label{DeltacalV}
\\
\Delta \widehat{\cal V}_0^{(F)} &\equiv&
            E\{- \delta^2 \widehat{{\cal G}}_{\bar{K}} \langle
(\hat{K}^{-1})^{{\bar{K}}L} \rangle
\delta^2 \widehat{{\cal G}}_L
+\delta^2 \widehat{{\cal G}}_{\bar{A}} \langle
(\hat{K}^{-1})^{{\bar{A}}B} \rangle
\delta^2 \widehat{{\cal G}}_B
\nonumber \\
&~&~~~ +\delta \widehat{{\cal G}}_{\bar{A}} \langle
       (\hat{K}^{-1})^{{\bar{A}}B} \rangle
\delta^{3'} \widehat{{\cal G}}_B + H.c.
\nonumber \\
&~&~~~ + \delta^2 \widehat{{\cal G}}_{\bar{\kappa}}
\langle (\hat{K}^{-1})^{{\bar{\kappa}}L}
 \rangle \biggl(\langle \hat{W}_{Lk} \rangle \delta \hat{z}^k +
{1 \over 2}\langle \hat{W}_{Lkl} \rangle \delta \hat{z}^{k}
\delta \hat{z}^{l} + \cdots \biggr)
\nonumber \\
&~&~~~ + H.c.\} ,
\label{DeltacalV(F)0}
\\
\Delta \widehat{\cal V}_0^{(D)} &\equiv&
    \langle Ref_{AB}^{-1} \rangle \langle \hat{D}^A \rangle
\langle \hat{K}_{{\hat{I}}\lambda} \rangle \delta
\hat{z}^{\hat{I}}
 (T^B)_\kappa^\lambda \delta \hat{z}^\kappa ,
\label{DeltacalV(D)0}
\\
\Delta \widehat{\cal V}_1^{(F)} &\equiv&
(m^{*'''}_{3/2})_{\bar{K}\hat{k}} \delta \hat{z}^{\hat{k}}
 \langle (\hat{K}^{-1})^{{\bar{K}}l} \rangle
\biggl(\frac{\partial \widehat{\cal W}_{eff}}
{\partial \hat{z}^l}-(m^{'''}_{3/2})_{lm}\delta \hat{z}^{m}
\nonumber \\
&~&~~~~ +(m_{3/2}+m_{3/2}^{''})_{l\bar{m}}\delta \hat{z}^{\bar{m}}
\biggr) +{\it H.c.}
\nonumber \\
&~& + E\{\delta \widehat{{\cal G}}_{\bar{K}} \langle
       (\hat{K}^{-1})^{{\bar{K}}B} \rangle
\delta^{3'} \widehat{{\cal G}}_B + H.c.
%\nonumber \\
+\delta \widehat{{\cal G}}_{\bar{\kappa}}
\delta^{2'} (\hat{K}^{-1})^{{\bar{\kappa}}\lambda}
\delta \widehat{{\cal G}}_{\lambda}
\nonumber \\
&~& +\delta \widehat{{\cal G}}_{\bar{\kappa}}
\delta (\hat{K}^{-1})^{{\bar{\kappa}}K}
 \delta^{2'} \widehat{{\cal G}}_K + H.c.\} ,
\label{DeltacalV(F)1}
\\
\Delta \widehat{\cal V}_1^{(D)} &\equiv&
    \langle Ref_{AB}^{-1} \rangle \langle \hat{D}^A \rangle
\langle \hat{K}_{{\hat{I}}{\hat{J}}\lambda} \rangle \delta
\hat{z}^{\hat{I}} \delta \hat{z}^{\hat{J}} (T^B)_\kappa^\lambda
\langle {z}^\kappa \rangle ,
\label{DeltacalV(D)1}
\\
\Delta {\cal V}'^{(F)} &=& E[Const.]^L
{1 \over 2} \langle \hat{W}_{Lkl} \rangle
\delta \hat{z}^{k} \delta
\hat{z}^{l} + H.c. ,
\label{DeltacalV'(F)}
\\
E[Const.]^L &\equiv& E [ \delta \hat{\cal G}_{\hat{I}}
 \delta(\hat{U})^{\hat{I}}_{\bar{I}}
\langle (\hat{K}^{-1})^{{\bar{I}}L} \rangle
 + \delta \hat{\cal G}_{\bar{I}} \langle
(\hat{K}^{-1})^{{\bar{I}}J} \rangle \delta(\hat{U})^{L}_J
\nonumber \\
&~& + \delta \hat{\cal G}_{\bar{I}}(\langle
(\hat{K}^{-1})^{{\hat{J}}L} \rangle
\delta(\hat{U}^{-1})_{\hat{J}}^{\bar{I}}
+ \langle (\hat{K}^{-1})^{{\hat{I}}{\bar{I}}} \rangle
\delta(\hat{U}^{-1})^L_{\hat{I}}) ].
\label{constL}
\end{eqnarray}
The quantities with a prime such as $\delta^{3'}
\widehat{{\cal G}}_B$ mean that the terms proportional to
$\delta^{2} \hat{z}^{\hat{I}}$ are omitted.
The ellipses in Eq.~(\ref{DeltacalV(F)0}) represent other terms
in $\delta^2 \widehat{\cal G}_K - \langle \hat{W}_{KL} \rangle
 \delta^2 \hat{z}^L$. (Refer Eq.~(\ref{gK2}).)
The soft SUSY breaking parameters $A$, $B_{\bar{k}}(z)_{eff}$
 and $C(\hat{z})_{eff}$ are given as
\begin{eqnarray}
A &=& m_{3/2}^{\ast '} - 3m_{3/2}^{\ast},
\label{Aagain}\\
B_{\bar{k}}(\hat{z})_{eff} &=& (m_{3/2}^{\ast}
+ m_{3/2}^{\ast ''}+m_{3/2}^{\ast '''})_{{\bar{k}}l} \delta
\hat{z}^l,
\label{Bkeff}\\
C(\hat{z})_{eff} &=&
E \delta \hat{\cal G}_{\bar{i}} \langle
(\hat{K}^{-1})^{\bar{i}j} \rangle
\biggl({1 \over 3!}\langle \hat{W}_{jIJJ'} \rangle \delta \hat{z}^I
\delta \hat{z}^J \delta \hat{z}^{J'}
+ {\langle \hat{W} \rangle \over M^2}
\delta^{2'}\hat{K}_j \biggr) + H.c.
\nonumber \\
&~&+ E\biggl(\delta \hat{\cal G}_{\bar{i}}
\delta^{2'}(\hat{K}^{-1})^{\bar{i}j}
\delta \hat{\cal G}_{j} + {\langle V \rangle \over M^2}
\delta^{2'} \hat{K} \biggr)
\nonumber \\
&~& - (m_{3/2}^{*'''})_{l\bar{l}} \langle
(\hat{K}^{-1})^{k\bar{l}} \rangle
(m_{3/2}^{'''})_{k{\bar{k}}} \delta \hat{z}^{\bar{k}} \delta
\hat{z}^l
\nonumber \\
&~& - (m_{3/2}^{'''})_{kl} \langle (\hat{K}^{-1})^{k\bar{l}} \rangle
(m_{3/2}^{*'''})_{\bar{k}\bar{l}} \delta \hat{z}^{\bar{k}} \delta
\hat{z}^l
\nonumber \\
&~& - \{ (m_{3/2}^{'''})_{ml}
\langle (\hat{K}^{-1})^{m\bar{k}} \rangle
(m_{3/2}^*+ m_{3/2}^{*''})_{{\bar{k}}k}
\delta \hat{z}^{k} \delta \hat{z}^{l} + H.c. \}
\nonumber \\
&~& + A \biggl[ E^{1/2}\biggl( {\langle \hat{W} \rangle \over M^2}
 \langle \hat{K}_{kl} \rangle
 - \langle \hat{K}_{kl\bar{i}} \rangle \langle
(\hat{K}^{-1})^{{\bar{i}}j} \rangle
 \delta \hat{\cal G}_j \biggr)
\nonumber \\
&~&~~~ + (m_{3/2}^{'''})_{kl} \biggr] \delta \hat{z}^k \delta
\hat{z}^l ,
\label{Ceff}
\end{eqnarray}
where
\begin{eqnarray}
(m_{3/2})_{k\bar{l}} &=& E^{1/2}{\langle \hat{W} \rangle \over M^2}
 \langle \hat{K}_{k{\bar{l}}} \rangle,
\label{mkl*}\\
m_{3/2}^{'} &=& E^{1/2}{\langle \hat{K}_{\bar{i}} \rangle \over M^2}
\langle (\hat{K}^{-1})^{{\bar{i}}j} \rangle \delta\hat{\cal G}_j,
\label{m'}\\
(m_{3/2}^{''})_{k\bar{l}} &=& -E^{1/2}
 \langle \hat{K}_{k\bar{l}\bar{i}} \rangle \langle
(\hat{K}^{-1})^{{\bar{i}}j} \rangle  \delta\hat{\cal G}_j,
\label{m''}\\
(m_{3/2}^{'''})_{\kappa\bar{l}} &=& -E^{1/2}
 \langle \hat{K}_{\kappa \bar{l}\bar{A}} \rangle \langle
(\hat{K}^{-1})^{{\bar{A}} \lambda} \rangle  \delta
\widehat{{\cal G}}_{\lambda},
\label{m'''kl*}\\
(m_{3/2}^{'''})_{\kappa l} &=& -E^{1/2}
 \langle \hat{K}_{\kappa l \bar{A}} \rangle \langle
(\hat{K}^{-1})^{{\bar{A}}\lambda} \rangle  \delta
\widehat{{\cal G}}_{\lambda}.
\label{m'''kl}
\end{eqnarray}
The $\Delta \hat{\cal V}^{(D)}_0$ and $\Delta \hat{\cal V}^{(D)}_1$
come from the $D$-term of the heavy gauge sector and are referred
to as the $D$-term contributions,
while the others are called the $F$-term contributions.

We should consider the renormalization effects for the soft SUSY
breaking parameters and diagonalize the scalar mass matrix
$V_{\hat{k}\hat{l}}$ to derive the weak scale SUSY spectrum,
which is expected to be measured in the near future.

\section{Features of the Effective Lagrangian}

The effective theory obtained in the previous section
has some excellent features.
We discuss two topics.

\subsection{Chirality Conserving Mass}
\label{subsec:mass-terms}

We discuss a {\em chirality-conserving} mass term
$(m^2)_{k\bar{l}}$, namely the
coefficient of $\delta \hat{z}^k \delta \hat{z}^{\bar{l}}$.
They are easily extracted from ${\cal V}_{Soft}^{eff}$ and
given by
\begin{eqnarray}
(m^2)_{k\bar{l}} &=& (m^2_0)_{k\bar{l}}
+ (\Delta \widehat{\cal V}_0)_{k\bar{l}}
+ (\Delta \widehat{\cal V}_1)_{k\bar{l}} ,
\label{m2_kl*}
\\
(m^2_0)_{k\bar{l}} &\equiv&
{\partial \over \partial \hat{z}^k}B_{\bar{m}}(\hat{z})_{eff}
\langle (\hat{K}^{-1})^{m\bar{m}} \rangle
 {\partial \over \partial \hat{z}^{\bar{l}}}B_{m}(\hat{z})_{eff}
\nonumber \\
&~&+ {\partial^2 \over \partial \hat{z}^k \partial
\hat{z}^{\bar{l}}} C(\hat{z})_{eff} ,
\label{m20_kl*}
\\
(\Delta \widehat{\cal V}_0)_{k\bar{l}} &\equiv&
{\partial^2 \over \partial \hat{z}^k \partial \hat{z}^{\bar{l}}}
\Delta \widehat{\cal V}_0^{(F)}
+ \langle Ref^{-1}_{AB} \rangle \langle \hat{D}^A \rangle
\langle \hat{K}_{{\bar{l}}\lambda} \rangle (T^B)^{\lambda}_k ,
\label{DeltaV0_kl*}
\\
(\Delta \widehat{\cal V}_1)_{k\bar{l}} &\equiv&
{\partial^2 \over \partial \hat{z}^k \partial \hat{z}^{\bar{l}}}
\Delta \widehat{\cal V}_1^{(F)}
+ 2 \langle Ref^{-1}_{AB} \rangle \langle \hat{D}^A \rangle
\langle \hat{K}_{k{\bar{l}}\lambda} \rangle
(T^B)^{\lambda}_{\kappa} \langle z^{\kappa} \rangle .
\label{DeltaV1_kl*}
\end{eqnarray}
The term $(m^2_0)_{k\bar{l}}$ is present before the
heavy sector is integrated out and so
it respects the original unified gauge symmetry.
On the other hand, other terms coming from
$\Delta \widehat{\cal V}$ can pick up effects of
the symmetry breaking.

The last terms in Eqs.~(\ref{DeltaV0_kl*}) and (\ref{DeltaV1_kl*})
 are the $D$-term contributions.
We discuss the conditions of those existence.
The non-zero VEV of the $D$-term is allowed
for a $U(1)$ factor, {\it i.e.} a diagonal generator from
the gauge invariance.
And the $D$-term for an unbroken generator cannot have its VEV.
Thus it can arise when the rank of the
gauge group is reduced by the gauge symmetry breaking.
The $D$-term contribution is proportional to the charge of
the broken $U(1)$ factor and gives mass splittings within
the same multiplet in the full theory.
We can rewrite $\delta^2 \hat{D}^A= \langle \hat{D}^A \rangle$ as
\begin{eqnarray}
\langle \hat{D}^A \rangle =
2(M_{V}^{-2})^{AB} E
 \delta \widehat{\cal G}_{\kappa} \delta \widehat{\cal G}
_{\bar{\lambda}}
 \{ G_{\bar{\mu}}^{\kappa\bar{\lambda}} (\hat{z} T^B)^{\bar{\mu}}
%\nonumber \\
   + G^{\bar{\mu}\kappa} (T^B)_{\bar{\mu}}^{\bar{\lambda}} \}
\label{<hatD>}
\end{eqnarray}
by using the gauge invariance.
We can see that the VEVs vanish up to $O(m_{3/2}^4/M_X^2)$
when the K\"ahler potential has the minimal structure.
Hence the sizable $D$-term contribution can appear only when
the K\"ahler potential has a non-minimal structure.

The other terms in Eqs.~(\ref{DeltaV0_kl*}) and
(\ref{DeltaV1_kl*}) are related to the $F$-terms.
%, so we call them $F$-term contributions.
They can be neglected in the case that the superpotential
couplings are weak and the $R$-parity conservation is assumed.
Therefore phenomenologically the $D$-term contribution
to the scalar masses is important to probe SUSY-GUT models
because it can give an additional contribution to squarks,
sleptons and Higgs bosons\cite{Sfermion}.

\subsection{Specific Case}

Finally we discuss the relation between
 our result and that in Ref.\cite{KMY2}.
For later convenience,
we list up features in the approach of Ref.\cite{KMY2}.
\begin{enumerate}

\item The starting theory is a unified theory obtained by
taking the flat limit of SUGRA with a certain type of total
K\"ahler potential,
so the terms of order $m_{3/2}^4(M_X/M)^n$ are neglected.
Since the unification scale $M_X$ is now believed to be lower
than the gravitational scale $M$ from LEP data\cite{LEP},
this procedure can be justified in such a model.
However it will be important when higher order corrections
are to be considered.
Then we must incorporate threshold effects and loop effects.

\item The scalar fields have canonical kinetic terms.
It was assumed that the SUSY fermion mass matrix and the
kinetic function can be diagonalized simultaneously, i.e.,
the relation $\langle \hat{K}_{\kappa{\bar{\lambda}}} \rangle
= \delta_{\kappa{\bar{\lambda}}}$ is imposed.

\item The {\it Hidden} assumption on superpotential was
taken because it was purposed to discuss consequences
independent of the details of each models.
The stability of gauge hierarchy is automatically guaranteed
under this assumption.

\item The heavy-light mixing, in general, can occur after soft
SUSY breaking terms are incorporated.
Then we must re-define the scalar fields by diagonalizing
the mass matrix.
It was assumed that there is no heavy-light mixing
after SUSY breaking.
\end{enumerate}

We shall derive the previous one ${\cal V}^{eff(non)}$ from
our scalar potential ${\cal V}^{eff}$
by refering to the list.
\begin{enumerate}
\item When we take the limit $M_X/M \longrightarrow 0$, we
 find that some terms vanish.
For example, $(m_{3/2}^{'''})_{\kappa\bar{l}}$ and
$(m_{3/2}^{'''})_{\kappa l}$ vanish
and
\begin{eqnarray}
\Delta \widehat{\cal V}_1 \longrightarrow
 E\{\delta \hat{{\cal G}}_{\bar{K}} \langle
 (\hat{K}^{-1})^{{\bar{K}}B} \rangle \delta^{3'} \hat{{\cal G}}_B
 + H.c.\}.
\end{eqnarray}

\item Next we impose the condition
$\langle \hat{K}_{\kappa{\bar{\lambda}}} \rangle
= \delta_{\kappa{\bar{\lambda}}}$.
Then $\Delta \widehat{\cal V}_1$ and some other terms vanish.
% under this condition and the limit $M_X/M \longrightarrow 0$.

\item Further we take the {\it hidden} ansatz
$\langle \hat{W}_{j...k...} \rangle = 0$.
Then the trilinear coupling constant is reduced to
\begin{eqnarray}
A E^{1/2} \langle \hat{W}_{klm} \rangle + {1 \over 2}
(m_{3/2}^* + m_{3/2}^{*''})_{k\bar{k}}
\langle (\hat{K}^{-1})^{{\bar{k}}n} \rangle \langle
\hat{W}_{nlm} \rangle.
\end{eqnarray}
%because $\langle \hat{W}_{jklm} \rangle = 0$.
%The gauge hierarchy problem is solved in this ansatz.

\item When we take a model with no heavy-light mixing,
$\Delta {\cal V}'^{(F)}$ does not exist.
We can find an ansatz for the K\"ahler potential
that the heavy-light mixing does not occur
in the gauge non-singlet sector after taking the flat limit.
For example, the ansatz
\begin{eqnarray}
K &=& K^{({\cal H})}(z^{{\cal H}}, z_{{\cal H}}^*
;\tilde{z}, \tilde{z}^*) +
K^{({k})}(z^{k}, z_{k}^*
;\tilde{z}, \tilde{z}^*)
\end{eqnarray}
fulfills our requirement.
\end{enumerate}

We find that ${\cal V}^{eff}$ reduces to $V^{eff(non)}$
after the above procedures.

\section{Conclusions}

We have derived the low-energy effective Lagrangian from
SUGRA with non-minimal structure and unified gauge symmetry
in model-independent manner.
The starting SUGRA is more general one than those considered
before.
The total K\"alher potential has a non-minimal structure
based on the hidden sector SUSY breaking scenario.
We have distinguished between the scales $M_X$ and $M$.

It is important to investigate its consequences at low-energy
because the non-minimal SUGRA appears naturally in many
circumstanses.
For example, SSTs lead to the non-minimal SUGRA effectively.
Even if SUGRA have the minimal structure at the tree level, it
can get renormalized and as a result, in general,
become non-minimal.

We have calculated the scalar potential by
taking the flat limit and integrating out the heavy sector.
The result is summarized in Eqs.~(\ref{calWeff})
--(\ref{m'''kl}).
We found new contributions to the soft terms
reflected to the non-minimality and the breaking of
unified gauge symmetry.
In particular, the sizable $D$-term contributions
generally exist in the scalar masses when the
rank of the gauge group is reduced by the gauge symmetry
breaking and the K\"alher potential has a non-minimal structure.
Its phenomenological implications were discussed
in Ref.\cite{Sfermion}.
Another important point is the gauge hierarchy problem.
Many SUSY-GUT models achieve the small
Higgs doublet masses by a fine-tuning of the parameters in the
superpotential.
If the SUSY breaking due to the hidden field condensations is
turned on, a SUSY breaking Higgs mass term can become heavy
 and the weak scale can be destabilized.
We have shown that the masses of light fields remain at the
weak scale if the couplings of hidden-sector fields to
visible-sector fields in the superpotential
satisfy certain requirements.

We have derived the results in Ref.\cite{KMY2} by taking
some limit and conditions.
We also have studied the SUGRA with Fayet-Iliopoulos $D$-term
 and derived the low-energy effective theory.

It is believed that the measurements of SUSY spectrum at
the weak scale can be useful in probing physics at SUSY-GUT
and/or SUGRA, if the SUSY breaking scenario through the
gauge-singlet sector in SUGRA is realized in nature.
Hence the precision measurements should be carried out by the
colliders in the near future.

\section*{Acknowledgements}
The author is grateful to H.~Murayama, H.~Nakano and I.~Joichi
and especially M.~Yamaguchi for useful discussions.
This work is supported by the Grant-in-Aid for
Scientific Research ($\sharp$07740212) from
the Japanese Ministry of Education, Science and Culture.

\appendix

\section{Consequences of $\langle \partial V /\partial
z^I \rangle = 0$}
\label{app:A}

In this appendix, we give some consequences of
the stationary conditions $\langle V_I \rangle \equiv
 \langle \partial V /\partial z^I \rangle =0$. From
Eq.~(\ref{V}), we find
\begin{eqnarray}
  V_I & =& M^{2}(e^{G/M^{2}})_I U +M^{2}e^{G/M^{2}} U_I
   +\frac{1}{2} (Re f^{-1})_{\alpha \beta, I}  D^\alpha D^\beta
\nonumber \\
  & & +(Re f^{-1})_{\alpha \beta} D^\alpha (D^\beta)_I
\nonumber \\
  &= &  G_I e^{G/M^{2}} U
    + M^{2}e^{G/M^{2}}\{ G_{IJ} (K^{-1})^J_{J'} G^{J'}
\nonumber \\
 &~& -G_{I'}(K^{-1})^{I'}_{J} K_{IJ'}^{J}(K^{-1})^{J'}_{I''}
G^{I''} +G_I \}
\nonumber \\
 & &  + \frac{1}{2} (Re f^{-1})_{\alpha \beta, I}
D^\alpha D^\beta + (Re f^{-1})_{\alpha \beta} D^\alpha
(z^{\dagger}T^\beta)_{J} K_{I}^{J}.
\label{VI}
\end{eqnarray}

Let us now multiply $(T^\alpha z)^I$ to the above, or project on
a heavy-real direction.
Using the identities derived from the gauge
invariance of the total K\"ahler potential
\begin{eqnarray}
   & & G_{IJ}(T^\alpha z)^J
   +G_J (T^\alpha )_I^J -K_I^J(z^{\dagger}T^\alpha)_J=0,
\label{G-inv1}
\\
   & & G_{IJJ'} (T^\alpha z)^{J'}
       +G_{IJ'} (T^\alpha)^{J'}_J
       +G_{JJ'} (T^\alpha)^{J'}_I
\nonumber \\
&~&~~~~~~~~~~~~~~~~~~~~~~~~~~~~~~-(z^{\dagger} T^\alpha )_{J'}
K_{IJ}^{J'}=0,
\label{G-inv2}
\\
   & & K_{IJ'}^{J} (T^\alpha z)^{J'}
       +K_{J'}^{J} (T^\alpha)^{J'}_I
       -[G^{J'} (z^{\dagger} T^\alpha )_{J'}]_{I}^{J}=0,
\label{G-inv3}
\end{eqnarray}
we obtain
\begin{eqnarray}
 V_I (T^\alpha z)^I &=&
       M^2 e^{G/M^2}(2+U/M^2) D^\alpha
       -F^I F^{*}_{J}(G^{I'}(z^{\dagger} T^\alpha )_{I'})_{I}^{J}
\nonumber \\
 & & + \frac{1}{2} (Re f^{-1})_{\beta \gamma, I} (T^\alpha z)^I
       D^\beta D^\gamma
\nonumber \\
 & &  + (Re f^{-1})_{\beta \gamma}
       (T^\alpha z)^I K_{I}^{J}(z^{\dagger}T^\gamma)_{J} D^\beta .
\label{pro-VI}
\end{eqnarray}
Taking its VEV, we find
\begin{eqnarray}
0&=& m_{3/2}^2 (2+\langle U \rangle /M^2) \langle
D^\alpha \rangle - \langle F^I \rangle \langle F^{*}_J \rangle
 \langle (G^{I'}(z^{\dagger} T^\alpha )_{I'})_{I}^{J} \rangle
\nonumber \\
 & & + \frac{1}{2}\langle (Re f^{-1})_{\beta \gamma , I}
(T^\alpha z)^I \rangle \langle D^\beta \rangle
\langle D^\gamma \rangle
\nonumber \\
&~&  +\frac{1}{2}\langle (Re f^{-1})_{\beta \gamma} \rangle
(M_{V}^2)^{\alpha \gamma} \langle  D^\beta \rangle,
\label{pro-<VI>}
\end{eqnarray}
where $(M_{V}^2)^{\alpha \beta}= 2\langle (z^{\dagger} T^\beta)_{J}
 K_{I}^{J} (T^\alpha z)^I \rangle$ is, up to the
normalization due to the gauge coupling constants,
the mass matrix of the gauge bosons.
Recalling that $(M_{V}^2)^{\alpha \beta}$ are assumed
to be $O(M_X^2)$ for broken generators of the GUT symmetry,
we conclude
\begin{eqnarray}
  \langle D^\alpha \rangle \leq O(m_{3/2}^2) ,
\label{<D>neq}
\end{eqnarray}
as the first three terms of Eq.~(\ref{pro-<VI>}) are already of
order $m_{3/2}^2 M_X^2$ or less.
It is noteworthy that quite a similar equation
to (\ref{<D>neq}) is obtained for the case of a  non-linear
realization of the gauge symmetry.

{}From Eqs.~(\ref{D}) and (\ref{<D>neq}), we find
\begin{eqnarray}
  \langle G^A \rangle \leq O(m_{3/2}^2/M_X) .
\label{<GA>neq}
\end{eqnarray}
By using the relations (\ref{GI}), we find
\begin{eqnarray}
&~& \langle \tilde{F}^*_i \rangle = O(m_{3/2} M),
	\label{<Fi>neq}\\
&~& \langle F^*_\kappa \rangle \leq O(m_{3/2} M_X).
	\label{<Fkappa>neq}
\end{eqnarray}

We now return to $\langle V_I \rangle=0$ itself.
Taking the VEV of Eq.~(\ref{VI}) and using the relations
(\ref{GI}) and (\ref{<D>neq}), we find
\begin{eqnarray}
&~& \langle M e^{G/2M^{2}} G_{Ij} \rangle \langle F^I \rangle
= O(m_{3/2}^2 M),\\
\label{Vi}
&~& \langle M e^{G/2M^{2}} G_{I\lambda} \rangle \langle F^I \rangle
\leq O(m_{3/2}^2 M_X).
\label{Vlambda}
\end{eqnarray}
Since $\langle M e^{G/2M^{2}} G_{IJ} \rangle
=\mu_{IJ}+O(m_{3/2})$,
the above reads
\begin{eqnarray}
&~&\mu_{Ij} \langle F^I \rangle = O(m_{3/2}^2 M),
\label{Viagain}\\
&~&\mu_{I\lambda} \langle F^I \rangle \leq O(m_{3/2}^2 M_X).
\label{Vlambdaagain}
\end{eqnarray}
Since we assume that $\mu_{KL}$ is $O(M_X)$
for heavy complex fields,\footnote{
Note that a careful analysis tells us that
$\langle F^K \rangle \leq O(m_{3/2}^2 \langle z^{K} \rangle /
M_K)$, where $M_K$ is the mass of $z^K$ from the
superpotential.
 Thus as far as  $\langle z^K \rangle \sim M_K $,  the VEV of
its $F$-term is always small $\sim m_{3/2}^2$.}
we find
\begin{eqnarray}
\langle F^K \rangle \leq O(m_{3/2}^2).
\label{<FK>neq}
\end{eqnarray}
%On the other hand,
The relation $\mu_{ij} = O(m_{3/2})$
%and $\mu_{i\lambda} = O(m_{3/2}M_X/M)$
is derived from the above relations (\ref{<Fi>neq})
and (\ref{Viagain}).
Therefore, in our convention, the
hidden-sector fields are contained in the light sector.

We can derive the following formula for the $D$-term condensation
\begin{eqnarray}
\langle D^\alpha \rangle &=&
2(M_{V}^{-2})^{\alpha \beta} \langle F^I \rangle
\langle F^{*}_J \rangle
\langle (G^{I'}(z^{\dagger} T^\beta)_{I'})_{I}^{J} \rangle
\nonumber \\
 &=& 2(M_{V}^{-2})^{\alpha \beta} \langle F^I \rangle
\langle F^{*}_J \rangle
\{ \langle G^{I'J}_I (z^{\dagger} T^\beta)_{I'} \rangle
\nonumber \\
&~&    + \langle G^{I'}_I \rangle (T^\beta)_{I'}^{J} \}
\label{D-formula}
\end{eqnarray}
from Eq.(\ref{pro-<VI>}).
We shall discuss the condition that sizable $D$-term
condensations of $O(m_{3/2}^2)$ exist.
In the case with minimal K\"alher potential,
 the formula (\ref{D-formula}) turns into simpler form as
\begin{eqnarray}
\langle D^\alpha \rangle
 &=& 2(M_{V}^{-2})^{\alpha \beta} \langle F^I \rangle
\langle F^{*}_J \rangle
          (T^\beta)_{I}^{J}.
\label{Min-D-formula}
\end{eqnarray}
It is shown that the $\langle D^{\alpha} \rangle$ is estimated
as less than $O(m_{3/2}^4/M_X^2)$ because $\langle F^K \rangle
\leq O(m_{3/2}^2)$ and $\langle F^A \rangle
= m_{3/2} \langle G^A \rangle
\leq O(m_{3/2}^3/M_X)$.
This result also holds in the presence of
Fayet-Iliopoulos $D$-term.
Hence we find that the existence of non-minimal K\"alher potential
is essential to the appearance of sizable $D$-term condensations.

\section{SUGRA with Fayet-Iliopoulos $D$-term}
\label{app:B}

In this appendix, we investigate the low-energy theory
derived from SUGRA with Fayet-Iliopoulos $D$-term.
This subject has not been completely examined in the literatures
\cite{N}\cite{CD}.
At first, we give comments about the SUGRA with Fayet-Iliopoulos
$D$-term.
(1) It necessarily has local $U(1)_R$ symmetry\cite{BFNS}.
It is expected that this symmetry become a key to solve
the problem of baryon and lepton number violation\cite{Fayet}.
On the other hand, the anomaly cancellation condition can
give a strong constraint on a model building\cite{CD}.
(2) Fayet-Iliopoulos $D$-term\cite{FI} is generated by
one-loop effects in SSTs with anomalous $U(1)$ symmetry\cite{DSW}.
The anomalies are cancelled by Green-Schwarz mechanism\cite{GS}.
Hereafter we denote Fayet-Iliopoulos $U(1)$ symmetry as $U(1)_R$.

Let us explain our starting point.
The gauge group is $G=G_{SM} \times U(1)_R$ where
$G_{SM}$ is a standard model gauge group
$SU(3)_C \times SU(2)_L \times U(1)_Y$.
Two types of chiral multiples exist.
One is a set of $G_{SM}$ singlet fields denoted as $\tilde{z}^i$.
Some of them have non-zero $U(1)_R$ charge and
induce to the $U(1)_R$ breaking.
We assume that the SUSY is broken by the $F$-term condensations of
 chargeless $\tilde{z}$'s.
The second one is a set of $G_{SM}$ non-singlet
fields $z^{\kappa}$.
For simplicity, we treat all $z^{\kappa}$'s as light fields.
Of course, we can generalize the case that the gauge group
is $G = G_U \times U(1)_R$ where $G_U$ is a unified group.

The scalar potential is given as
\begin{eqnarray}
   V &=& V^{(F)} + V^{(D)} ,
\label{VFI}
\\
   V^{(F)} &\equiv& M^{2}exp(G/M^{2})
(G^I (G^{-1})_I^J G_{J}-3M^{2}) ,
\label{V(F)FI}\\
V^{(D)} &\equiv& \frac{1}{2} (Re f^{-1})_{\alpha \beta}
 D^{\alpha} D^{\beta} + \frac{1}{2} (D^{R})^2 ,
\label{V(D)FI}
\end{eqnarray}
where the index $I$, $J$,... run all scalar species and
$D^R \equiv g_R G_I (Q^R z)^I$.
Here we denote the gauge coupling constant and
$U(1)$ charge of $U(1)_R$ as $g_R$ and $Q_R$, respectively.
We find that $D^R$ contains a constant term $2g_R M^2$ since
\begin{eqnarray}
 D^R = g_R (K_I (Q^R z)^I + 2 M^2) ,
\label{DR}
\end{eqnarray}
where we used the fact that the superpotential carries
$U(1)_R$ charge 2, i.e.,
\begin{eqnarray}
\frac{\partial W_{SG}}{\partial z}(Q^R z)^I
         = 2 W_{SG}.
\label{G-sym}
\end{eqnarray}
It is easy to find that Fayet-Iliopoulos $D$-term\cite{FI} exists
in the second term of $V^{(D)}$.
Note that the coefficient of Fayet-Iliopoulos $D$-term is fixed
from the $U(1)_R$ symmetry and
this fact is essential to the conclusion that no sizable
$D$-term contribution to scalar masses appears in the SUGRA
with the minimal K\"ahler potential.

The $U(1)_R$ is broken by the condensations of $\tilde{z}$
because $V^{(D)}$ is a dominant part in $V$.
The orders of those VEVs are estimated as
$\langle \tilde{z} \rangle = O(M)$.
Hence the breaking scale of $U(1)_R$ is of order $M$.

Now we compute the scalar potential of the low-energy effective
theory by taking the flat limit
and integrating out the heavy fields in $\tilde{z}$'s
simultaneously.
%The procedure is almost same as done in subsection 3.3.
The $D$-term contribution is added to the scalar masses
in comparison with the result in Ref.\cite{S&W}.
We write it down in the form that the scalar masses are read off,
\begin{eqnarray}
V^{(FI)} &=& V_{SUSY}^{(FI)} + V_{Soft}^{(FI)} + \Delta V^{(FI)},
\label{V-FI}\\
V_{SUSY}^{(FI)} &=&
\frac{\partial \widehat{\cal W^*}}{\partial z^*_\kappa}
\langle (K^{-1})_\kappa^\lambda \rangle
\frac{\partial \widehat{\cal W}}{\partial z^\lambda}
 + {1 \over 2}g_\alpha^2 (\langle K_\kappa^\lambda \rangle
z_\lambda^* (T^\alpha)^\kappa_\mu z^\mu)^2,
\label{VSUSY-FI}\\
V_{Soft}^{(FI)}&=& A \widehat{W}
+ B^\kappa(z)
\langle (K^{-1})_\kappa^\lambda \rangle
\frac{\partial \widehat{\cal W}}{\partial z^\lambda}
		+ {\it H.c.}
\nonumber \\
&~&+ \{(|m_{3/2}|^2 + {\langle V \rangle \over M^2})
\langle K_\kappa^\lambda \rangle  + \langle \tilde{F}^i \rangle
(\hat{K}_{i\kappa}^{\mu} \langle (K^{-1})_{\mu}^{\nu} \rangle
\hat{K}^{j\lambda}_{\nu}
- \hat{K}_{i\kappa}^{j\lambda}) \langle \tilde{F}^{*}_j \rangle
\nonumber \\
&~&~~~~~ + g_R \langle D^R \rangle
Q_{\kappa}^R \langle K_\kappa^\lambda \rangle \}
z^\kappa z^*_\lambda
\nonumber \\
&~& +\{- \langle \tilde{F}^i \rangle H_{i\kappa\lambda}^j
\langle \tilde{F}^{*}_j \rangle
 + m_{3/2} \langle \tilde{F}^i \rangle H_{i\kappa\lambda}
\nonumber \\
&~&~~~~~ + m_{3/2}^* \langle \tilde{F}^*_j \rangle
H_{\kappa\lambda}^j \} z^\kappa z^\lambda + H.c. ,
\label{VSoft-FI}\\
\Delta V^{(FI)} &=&
\frac{\partial \widehat{W^*}}{\partial \tilde{z}^*_i}
\langle (K^{-1})_i^j \rangle
\frac{\partial \widehat{W}}{\partial \tilde{z}^j}
 + \langle \tilde{F}^i \rangle
\frac{\partial \widehat{W}}{\partial \tilde{z}^i} + H.c.
\label{DeltaV-FI}
\end{eqnarray}
up to constant terms and higher order terms of $O(m_{3/2}^5/M)$.
Here $B^\kappa (z)$, $C(z, z^*)$, $\widehat{\cal W}$
and $\widehat{W}$ have been already defined in subsection 2.3.

We find that the $U(1)_R$ $D$-term contribution to scalar masses
%(the last term in Eq.(\ref{VSoft-FI}))
which can destroy universality among scalar masses at $M$.
(Its existence was suggested in ref.\cite{N}, but we have proved
 it by deriving the full low-energy scalar potential from SUGRA
directly.)
As its contribution is proportional
to the $U(1)_R$ charge, the $U(1)_R$ charge of
 matters can be known from the measurements of weak scale SUSY
spectrum.

Our procedure and result are applicable to
the effective SUGRAs derived from SSTs.
It is known that there are many string models with $G \times
U(1)^n$\cite{IK} and some models generate Fayet-Iliopoulos
$D$-term\cite{C}.
Therefore it is important to search for a realistic string model
by taking care of extra $U(1)$ symmetries.

\end{document}